\documentclass[preprint2]{aastex702}

\newcommand{\reply}[1]{{\color{black}{#1}}} 
\newcommand{\replytwo}[1]{{\color{black}{#1}}} 
\newcommand{\replythree}[1]{{\color{black}{#1}}} 
\newcommand{\replyfour}[1]{{\color{black}{#1}}} 
\newcommand{\replyfive}[1]{{\color{black}{#1}}} 

\graphicspath{{./}{figures/}}
\usepackage{graphicx}
\usepackage{gensymb}

\begin{document}

\title{ A Search for Dust in the Outflow of the HH212 Class 0 Protostellar System}


\author[0000-0002-9912-5705]{Adele L.\ Plunkett}
\affiliation{U.S. National Science Foundation, Randolph Building, 401 Dulany St, Alexandria, VA, 22314, USA}
\affiliation{National Radio Astronomy Observatory, 520 Edgemont Rd, Charlottesville, VA, 22903, USA}
\email{aplunket@nrao.edu}

\author[0000-0002-7993-4118]{Per Bjerkeli}
\affiliation{Chalmers University of Technology, Department of Space, Earth and Environment, SE-412 96 Gothenburg, Sweden}
\email{per.bjerkeli@chalmers.se}

\author[0000-0002-3835-3990]{Jon P.\ Ramsey}
\affiliation{Department of Astronomy, University of Virginia, Charlottesville, VA, 22904, USA}
\email{jon.p.ramsey@gmail.com}

\author[0000-0001-6307-4195]{Daniel Harsono}
\affiliation{Institute of Astronomy, Department of Physics, National Tsing Hua University, Hsinchu, Taiwan}
\email{dharsono@gapp.nthu.edu.tw}

\author{Hannah Shoemaker}
\affiliation{Department of Astronomy, University of Virginia, Charlottesville, VA, 22904, USA}
\affiliation{Department of Physics, University of Virginia, Charlottesville, VA, 22904, USA}
\email{hannahjshoemaker@gmail.com}

\correspondingauthor{A.\ Plunkett}
\email{aplunket@nrao.edu}

\begin{abstract}

Perhaps the most visually striking phenomenon during star formation is the ejection of material along the direction of the protostar's rotational axis. Observed in the form of molecular outflows and/or atomic jets, these features carry away mass, energy and angular momentum from the system. Given the known dust content of protoplanetary disks, and the energetics of outflows, it has been suggested that it should be possible to lift dust from disks via winds, yet limited observational evidence has been presented in the literature. Any detection of dust within protostellar outflows not only puts constraints on the dust content in outflows, but if a sufficient amount of dust can be removed from the system,  the mass budget for planet formation could also be affected. We present re-imaging of archival, multi-frequency dust continuum observations taken with the Atacama Large Millimeter/sub-millimeter Array (ALMA) towards the HH212 protostellar system. We detect an extension in the dust emission, and using multi-frequency data in Bands 7 and 9, we estimate the spectral index of the emitting dust and the dust mass in the wind launching region. We find a small but still significant amount of dust mass coincident with the molecular outflow cavity. This finding is assessed in relation to the hypothesis that the outflow could be capable of transporting dust away from the HH212 protoplanetary disk.

\end{abstract}

\section{Introduction}
\label{sec:introduction}

Protostellar outflows are ubiquitously observed phenomena in regions of active star formation. Through the combination of the gravity due to a protostar, magnetic fields, and the centripetal acceleration experienced by gas orbiting the protostar, material and angular momentum can be ejected from the disk along magnetic field lines \citep{BlandfordPayne1982,PudritzNorman1983_diskwinds}. Although the primary outcome of protostellar outflows is the efficient removal of angular momentum from protoplanetary disks, they also remove some mass in form of dust and gas from the system. The details of how much mass is removed depends on the adopted wind-launching model and the local physical conditions \citep[e.g.][]{Sheikhnezamietal2012_diskwinds,Gresseletal2015_diskwinds,GresselRamseyetal2020_diskwinds}. 

Indeed, the launching of a protostellar outflow and subsequent removal of mass and angular momentum could have a significant impact on the structure of the disk, particularly within the outflow launching region. This region is also particularly important for rocky planet formation.  Since planets necessarily form out of disk material, if the disk structure can be significantly affected by outflows, there could also be a significant impact on the initial conditions for planet formation \citep[e.g.,][]{Bai:2017qf,Pascucci2023}. That said, the details of the wind-launching process are still debated \citep[e.g.][]{Franketal2014_PPVI}. Several theories for magnetically-driven wind launching have been proposed, and while the physics are very similar, they differ mainly in where the launching takes place. The first is that of a stellar wind that emanates from the poles of the protostar \citep[e.g.][]{Bouvier14}. The second is a so-called `X-wind', where an asymptotically radial outflow is launched at the location where the magnetic field of the disk interacts with the protostellar magnetic fields (the so-called `X-point'; e.g.\ \citealt{Shu2000,Shang:2006lr}). The third, and currently preferred theory, is that of a magnetic disk wind emanating from an extended region of the surface of a protoplanetary accretion disk \citep{BlandfordPayne1982,PudritzNorman1983_diskwinds}. Several studies in recent years indeed indicate that extended disk winds are common during the early stages of protoplanetary evolution \citep[e.g.][]{bjerkeli2016,Taboneetal2017_SO_SO2,Lee2018b,devalonetal2020_dgtaub}. In reality, however, it is likely that a combination of these launching mechanisms will be required to understand all of the observations \citep[e.g.][]{Shang:2006lr,Banerjee:2006lr,Franketal2014_PPVI,RamseyClarke2019_diskwinds}.

Outflows have historically been probed through molecular and atomic emission lines across a range of wavelengths (and energy regimes) \citep[e.g.][]{Snelletal80,Hartiganetal2005,Podioetal2006, Bjerkelietal2012,Plunkettetal2015,Liuetal2016, Kristensenetal2017}. However, due to insufficient spatial resolution, until recently, probing the launching region ($< 10$ au) directly has been problematic. This situation dramatically changed with the construction of the Atacama Large Millimeter/sub-millimeter Array (ALMA), where sensitivity and angular resolution permit observations on scales less than 10 au at sub-millimeter wavelengths in nearby star formation regions. The high angular resolution and excellent sensitivity of ALMA also opens up the possibility to study the dust content in the outflow launching region simultaneously with the molecular emission. 

If a disk wind is responsible for accelerating gas from the accretion disk into the outflow, it is conceivable that small dust particles can also be swept up and carried away. If a sufficient mass of cool, millimeter-size dust is entrained in the outflow, it should be observable, and maybe even spatially resolved, with ALMA.  Some evidence of asymmetric dust structures spatially coincident with outflow cavities in young protostellar sources has been presented previously on scales of $\sim10-10^3$\ au \citep{maury2018,Legouellec2020,Sabatini2025}, and on larger scales there have been reports of continuum emission tracing knots along cavity walls \citep[e.g.][]{maury2012}. However, it was not until the recent result by \citet{Sabatini2025} that observations of such a system -- in their case, the Class I source L1551 IRS5 -- sufficiently probed inner structures close to the disk, spatially resolved dust spectral index values, and therefore enabled discussion of dust characteristics relative to cavity structures. It is important to point out that dust continuum emission carries no kinematic information. As a result, spatially resolved observations of any dust continuum needs to be compared with spatially and spectrally resolved molecular line observations in order to constrain the true origin of the dust continuum emission.

To discuss the various components of a complex protostellar system, it will be useful to define the terminology we use herein. In relation to protostellar systems, the term `outflow' is commonly used to describe all material leaving the system, including any material that is being swept up during the launching process. Within an outflow, there are several distinct components. The jet is defined as the launched atomic and molecular material which becomes well-collimated and attains a high velocity ($\gtrsim$30 km/s). The wind is defined as the mostly molecular material that is launched at lower velocities and is typically less collimated than the jet. The cavity wall is defined as the boundary region between the molecular wind and the ambient envelope material, encompassing the outflow. Material that is being entrained by the wind is generally accepted to originate from the region close to the cavity wall. As defined here, it contains envelope or ambient material that has been dragged along by the molecular wind. In contrast, we define launched material as the gas and dust that is ejected or launched from the disk and/or protostar. Meanwhile, we refer to ``dust in the wind'' as dust continuum emission close to the disk -- in proximity to the cavity wall, within a few au of the disk -- but that is not co-spatial with the disk or the infall region of the envelope where matter is falling inwards. We acknowledge that -- with finite angular resolution -- it is difficult to define a sharp boundary between where we observe entrained ambient material versus dust that has been lifted directly from the disk. However, any dust continuum emission in the region close to the disk (but outside of it), must either be moving outwards, or falling inwards. As such, even the entrained dust close to the cavity wall could originate in the disk or the innermost infalling envelope.

In this paper, we present spatially resolved archival continuum data from ALMA that suggests the presence of dust in the disk wind of the young protostellar system HH212. Based on a systematic archival search (Sect.\ \ref{sub:search}), we identified a handful of sources potentially showing evidence of thermal dust emission that extends co-spatially from the base of the outflow region.  \replyfour{Due to a combination of source geometry, well-characterized gas kinematics, and the multitude of multi-band observations available for public use in the ALMA Science Archive (ASA), HH212 was selected} as a unique target for our scientific goals.

\replyfour{HH212} is a Class 0 protostellar system located at $\sim$400pc in the L1630 cloud of Orion \citep{Kounkel17}.  The protoplanetary disk of HH212 is nearly edge-on with inclination of -4$^{\circ}$, it is oriented with a position angle of -23$^{\circ}$ \citep{Claussen1998}, and the central protostellar mass is estimated to be $0.25\pm 0.05$ $M_\odot$ \citep{Lee2017SciAdv}.  \replyfour{Numerous investigations of molecular line emission have been undertaken to characterize the gas dynamics around HH212.  Through these works, a resolved dusty disk, a highly collimated rotating jet, and a slower structure composed a wide angle wind and cavity have been traced by species such as SO, SO$_2$, C$^{34}$S, and CO \citep{Codellaetal2014_HH212,Lee2014,Podio2015_SO,Lee2017NatAs,Taboneetal2017_SO_SO2,Lee_2018a,Lee2018b,Lee_2019,Tabone20,LeeLietal2022,Lopezvazquez2024}. Observations have also revealed the flattened and rotating envelope near the disk, which makes the origin of emission outside of the disk ambiguous, with possible contributions from both envelope and outflow.
\citep[e.g.][]{Codellaetal2014_HH212,Lee2017ApJ,Bianchietal2017_HH212methanol,LeeTaboneetal2021,LeeLietal2021b}.}



\replyfour{It remains unresolved whether hints of extended dust emission are associated with the wind-launching region, and HH212 is an excellent case study for this exploration.  Our paper provides multi-band, optimized imaging of this source to constrain the morpholgy, spectral index, and mass estimates of dust in the region of the protostellar system corresponding to the cavity.}  \replyfour{The paper is organized as follows.  We describe the archival data and data processing in \S \ref{sec:data}.  In \S \ref{sec:results}, we present the emission maps (\S \ref{sub:emission}), spectral index map (\S \ref{sub:spectralindex}), and dust masses in distinct regions with respect to the disk and cavity (\S \ref{sub:dustmass} ).  We discuss the implications in \S \ref{sec:interpretation}, and provide a summary in \S \ref{sec:conclude}.}

\section{Data}
\label{sec:data}
\subsection{Archival search and target selection}
\label{sub:search}
The ASA contains data from all observations that have been completed and delivered by ALMA, and the majority of these data are publicly available. We conducted a programmatic search of the ASA to find sources with the science category \replyfour{``ISM and star formation,'' and with science keywords ``outflows, jets and ionizing winds,'' ``low mass star formation,'' or ``high mass star formation.''} Since we wanted to investigate the regions near protoplanetary disks in detail, and in particular the regions where winds are launched \citep[e.g.][]{Franketal2014_PPVI,bjerkeli2016,Lee2017NatAs}, and where we believe it is most likely to find dust in the wind, we further refined our search by filtering for observations with an angular resolution of $<$ 0.1 arcseconds; this corresponds to better than about 100-400 au in nearby, low-mass protostars. Finally, based on manual inspection of the data and the literature, only sources with known outflows and protoplanetary disks , that also have been previously observed with sufficient angular resolution, were retained for deeper investigation and analysis.

As a final step, we further refined the search by selecting only sources which had continuum emission that was significantly non-Gaussian (i.e.\ that deviates from the two-dimensional Gaussian shape that is typically used to describe protoplanetary disks), plus a comprehensive set of archival data products across multiple frequencies (hence, observed with different ALMA bands). Ultimately, this search permitted us to identify one source with apparent continuum emission in the wind-launching region and a significant amount of publicly available data in the ASA: HH212.

\subsection{Re-imaging of archival data}
\label{sub:reimaging}
The ASA currently contains public, high angular resolution data for HH212 in ALMA Bands 3, 6, 7 and 9. This allows for the calculation of the spectral index, which can be used to put constraints on the size of the dust (see \ref{sub:spectralindex}). Nine archival data sets were selected, and are listed in Table \ref{tab:observational_parameters} (Appendix \ref{appendix:band7}). For each data set, the calibrated measurements sets (MSs) were requested via the ALMA Regional Centers and produced following the standard pipeline. Using the calibrated MSs, we performed re-imaging with the Common Astronomy Software Application (CASA), version 5.6.1 \citep{McMullingetal2007_CASA}, for each corresponding ALMA band with intent to recover faint emission, in particular, beyond the known protoplanetary disk.  We note that, for the Band 9 observations only, phase-based self-calibration was performed to enhance the signal-to-noise, and then re-imaged. 

\begin{figure}[ht]
\centering
\includegraphics[width=\columnwidth]{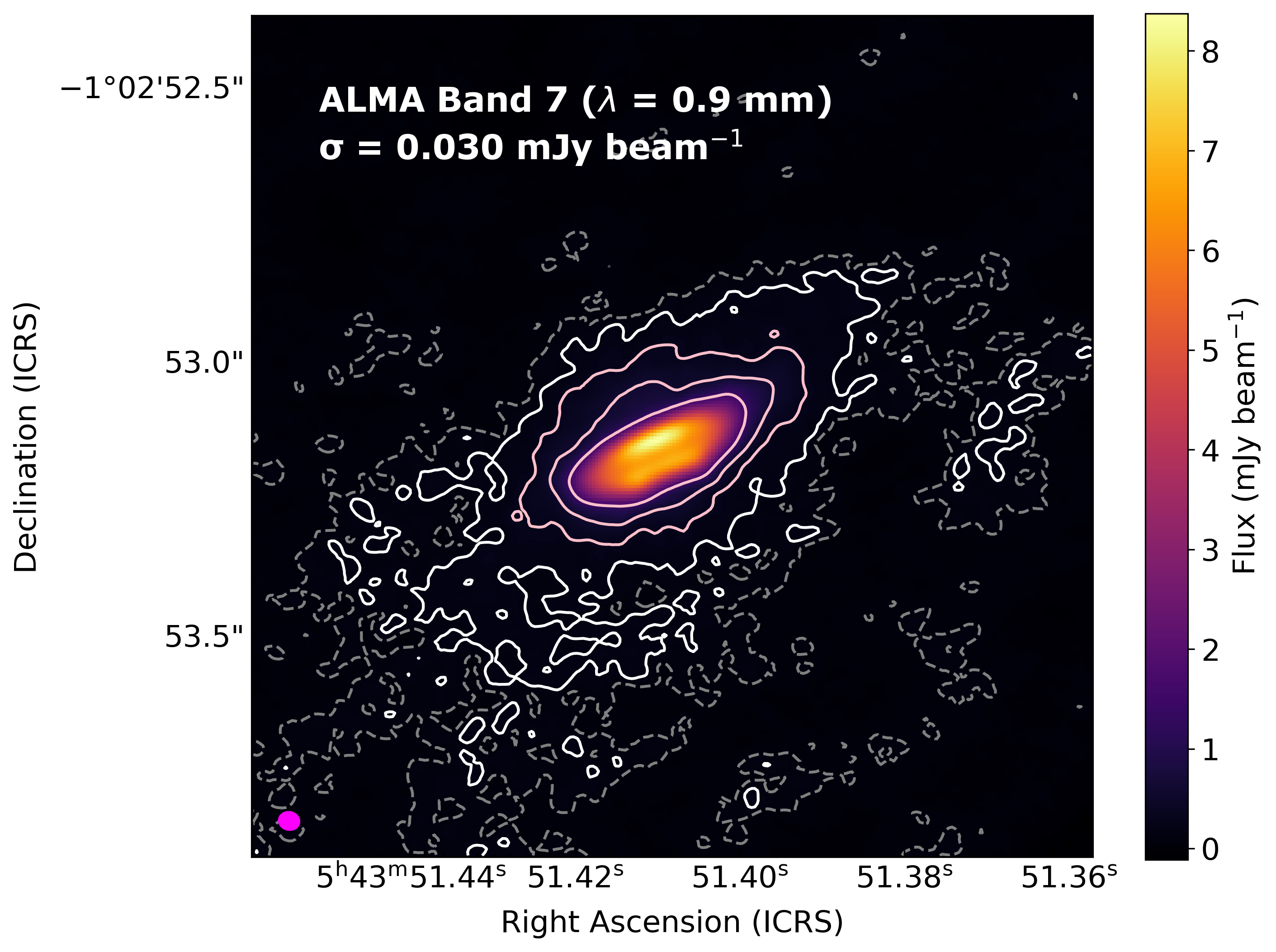}
\caption{Band 7 image of dust continuum emission towards HH212, combining data from four ALMA projects. The outermost (gray dashed) contour line denotes 3$\sigma$, the white contour marks 5$\sigma$, and the other (pink) contours are at 10$\sigma$, 20$\sigma$ and 50$\sigma$. Image parameters are given in Table \ref{tab:dataimageparams}.}
\label{fig:B7continuum}
\end{figure}

In the case of the Band 7 data, we combined data sets across different projects (Table \ref{tab:observational_parameters}) to obtain a high angular resolution, high-fidelity Band 7 continuum image; the details of the resulting image properties after cleaning can be found in Table \ref{tab:band7combination} (Appendix \ref{appendix:band7}). There were a total of 76 spectral windows with continuum emission across the Band 7 projects. As an intermediate step, we imaged the data from each project separately using the spectral mode ``cube'' of the \texttt{tclean} task in CASA; we manually searched for any channels displaying significant line emission so that we could ensure that the combined continuum image only included emission-free channels.  This procedure also allowed an opportunity to identify any potential data quality issues that could have affected a single dataset, and ultimately no significant issues were found. For a final image, the projects were combined using CASA \texttt{tclean} with Briggs weighting (\texttt{robust} = 0.5, which balances between sensitivity and angular resolution). We invoked the ``interactive'' parameter in order to manually adjust the clean mask at certain intervals. The resulting Band 7 image is presented in Figure \ref{fig:B7continuum}.  The same imaging procedure and parameters were subsequently used to obtain images in Bands 3, 6 and 9. Project IDs and their observational set ups are listed in Table \ref{tab:observational_parameters}.

\begin{figure*}[ht]
\centering
\includegraphics[width=\textwidth]{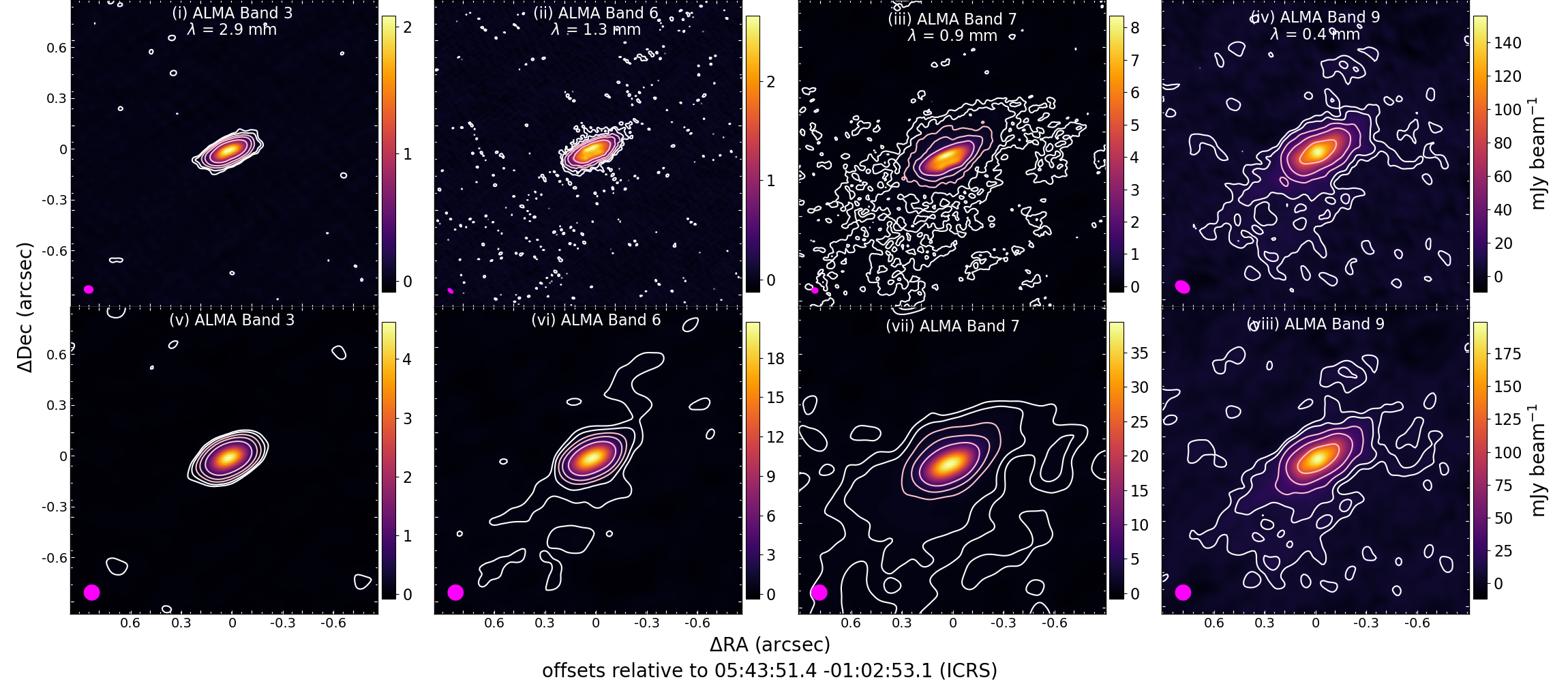}
\caption{Continuum emission towards HH212 at multiple frequencies. The \emph{top} row of the figure presents images across ALMA Bands at their native cell and beam size. The \emph{bottom} row presents the same data, but convolved to the same beam and cell size based on the Band 9 data as a template. Contours are drawn at 3$\sigma$ and 5$\sigma$ (white), and 10$\sigma$, 20$\sigma$, and 50$\sigma$ (pink) for each image (see Table \ref{tab:dataimageparams}). \replyfour{The scaling in each panel is according to the peak intensity of each image, and colorbars are shown (in units of mJy/beam) for each panel.} More information about image-specific parameters can be found in Table \ref{tab:dataimageparams}.}
\label{fig:allbands_cont}
\end{figure*}

The resulting images for each band are presented in Section \ref{sub:emission} and Figure \ref{fig:allbands_cont}, while resulting beam sizes and root mean square (rms) noise levels are given in Table \ref{tab:dataimageparams}. Images which are indicated as ``smoothed'' were convolved to a common beam size of 0.09$\arcsec$ using CASA task \texttt{imsmooth}. The different images were further re-gridded using CASA task \texttt{imregrid} as needed to ensure they have the same cell size (0.016$\arcsec$) as the Band 9 image. \replyfour{A common beam size and gridding are needed for cross-band comparison and pixel-by-pixel spectral index calculation, which are  described in \S \ref{sub:spectralindex}.  The coarsest resolution determines the common beam size, which in this case is given by the Band 9 dataset to be 0.09$\arcsec$.  For all bands, we smoothed to a round beam size of $0.09\arcsec\times0.09\arcsec$ for simplicity and to minimize any geometric artifacts related to the beam shape.  The sensitivity of the images, on the other hand, still differs among the images in different bands after smoothing.  We discuss the implications in \S \ref{sub:emission}.}

\begin{table*}[ht]
\setlength{\tabcolsep}{3pt} 
  \begin{center}
    \caption{Image parameters resulting from the re-imaging process}
    \label{tab:dataimageparams}
    \begin{tabular}{lcccccccc}
    \hline
    \hline
      \textbf{Band} &\textbf{Smoothed} &  \replyfive{\textbf{$\theta_{beam}$\ (Maj.$\times$Min.) }} & \textbf{P.A.} & \textbf{Cell Size} & \textbf{$\sigma_{RMS}$} & \textbf{Peak Intensity}& \textbf{Total Flux} \\ %
    & & [\arcsec] & [$^{\circ}$] & [\arcsec] & [mJy beam$^{-1}$]& [mJy beam$^{-1}$] & [Jy] \\
     \hline
    3 & No & 0.053 x 0.045 & 81.70 & 0.005 & 0.011 &   2.087 & 0.013 \\
    6 & No & 0.036 x 0.019 & 48.03 & 0.005 & 0.023 &   2.664 & 0.061 \\
    7 & No & 0.038 x 0.033 & 77.31 & 0.006 & 0.030 &   8.375 & 0.194 \\
    9 & No & 0.090 x 0.066 & 56.57 & 0.012 & 1.859 & 156.220 & 1.220 \\
      \hline
    3 & Yes & 0.090 x 0.090 & -- & 0.012 & 0.015 &   4.625 & 0.011 \\
    6 & Yes & 0.090 x 0.090 & -- & 0.012 & 0.103 &  20.702 & 0.065 \\
    7 & Yes & 0.090 x 0.090 & -- & 0.012 & 0.140 &  39.417 & 0.214 \\
    9 & Yes & 0.090 x 0.090 & -- & 0.012 & 2.198 & 198.692 & 1.304 \\

      \hline
    \end{tabular}
    \tablecomments{\replyfive{Synthesized beamsize ($\theta_{beam}$) is given in terms of major and minor axes (Maj.$\times$Min.) in arcseconds, and position angle (P.A.) in degrees.} Reported total flux is calculated after masking pixels below 5$\sigma$ for a given image.}
  \end{center}
\end{table*}

\section{Results}
\label{sec:results}

\subsection{Emission maps and dust morphology}
\label{sub:emission}

As mentioned, the disk of HH212 is oriented roughly edge-on, with the major axis along the northwest-southeast direction. Maps of the continuum emission detected in each ALMA band are shown in Figure \ref{fig:allbands_cont}. The top panel of Figure \ref{fig:allbands_cont} shows the images at their native resolution \replyfour{to preserve the highest-resolution (available) morphology for each band}, while the lower panel shows the same images \replyfour{but with a common angular resolution of 0.09\arcsec\ to reveal comparable structures across bands.}

The shapes of the observed continuum in ALMA Bands 3 and 6 are very similar to Gaussians \replyfour{down to the $3\sigma$\ and $5\sigma$ noise levels, respectively}. The images for ALMA Bands 7 and 9, meanwhile, tell a different story. The continuum emission in these bands show significant departures from a Gaussian distribution in the south-southwest direction, in the same direction as the red-shifted outflow \citep[e.g.][]{Taboneetal2017_SO_SO2}. This extension is apparent in the images at both native resolution and when convolved to a common beam size. The extension in Band 7 can be more clearly seen at greater than a $3\sigma$ level in Figure \ref{fig:B7continuum}, and non-Gaussianity is apparent up to a level of at least $10\sigma$. Although Band 7 was observed at higher angular resolution and to better sensitivity than the Band 9 data, the extension is still apparent in Band 9 and above the $3\sigma$ level.  Relative to the Band 7 image, the extended emission in Band 9 \replyfour{appears to break} into two branches, one extending south-southwest of the source, approximately perpendicular to the disk major axis, and the other extending at least 0.5\arcsec south-east of the disk roughly parallel to the disk major axis, but with a significant thickness in the southern direction. \replyfour{We note that a direct comparison of low-level extended morphology is limited by the differing rms noise levels between the Bands 7 and 9 data, and the exact limits of the $3\sigma$\ and $5\sigma$ contour levels should be interpreted cautiously.  Further, even though the rms noise of the Band 9 data seems to be worse than those of the other bands, in terms of sensitivity to dust mass, Bands 7 and 9 would be the most sensitive.  See \ref{sub:dustmass} for more details.}

\begin{figure*}[ht]
\centering
\includegraphics[width=0.9\textwidth]{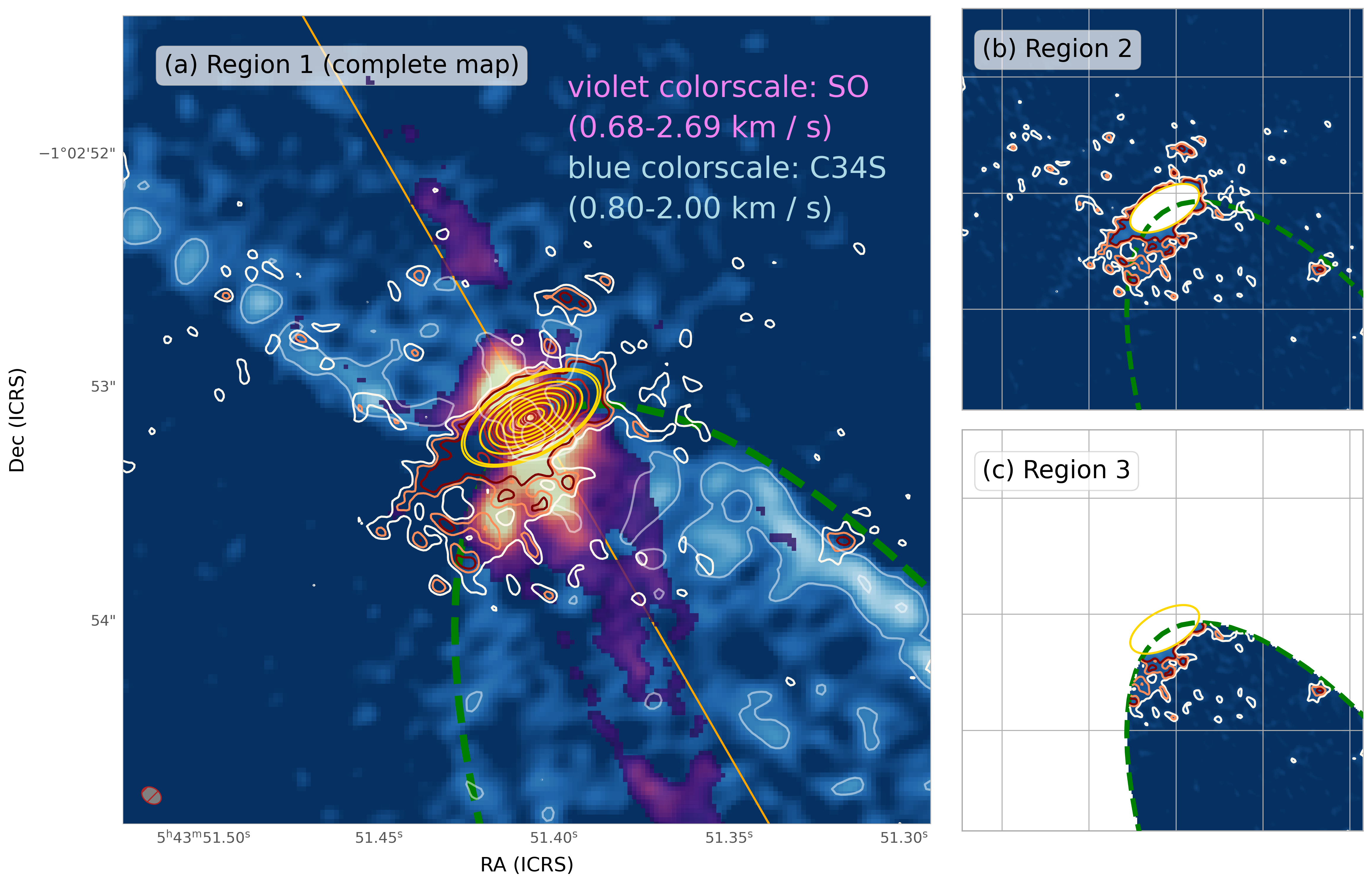}
\caption{
\replyfour{(a)} SO (9$_8$-8$_7$) and C$^{34}$S (7-6) integrated emission moment 0 maps using data from \cite{Taboneetal2017_SO_SO2} (violet, blue color scales, respectively); SO is masked below 2.5 times the rms noise level of the moment map, and the light blue line contour signifies the $3\sigma$ level of the C$^{34}$S moment map. (See Figure \ref{fig:appendix_soso2} for more detailed view of specific molecular lines, including C$^{17}$O.) Overlaid is the re-imaged Band 9 continuum emission (white, salmon, and red line contours at 3, 4, and $\geq 5\sigma$, respectively). The dashed green \replyfour{ellipse} encloses the region where significant C$^{34}$S and SO emission is observed to the south-southwest, which we adopt as the breadth of the outflow cavity on the red-shifted side of the outflow. The yellow contours show a two-dimensional Gaussian fit to the central emission in Band 9 (and shown at intervals of 4$\sigma$, 5$\sigma$, 10$\sigma$, 15$\sigma$, etc., where $\sigma$ is the rms value of the Band 9 data). The orange line denotes the bisection of C$^{34}$S emission lobes, and the beam size is indicated at the bottom left \replyfour{of panel (a).  Panel (a) represents what we refer to as ``Region 1,'' with no masking for the purpose of dust mass calculations.  Panels (b) and (c) show continuum emission only in masked regions used for mass estimates outside the disk (Region 2) and inside the cavity (Region 3), respectively (see Appendix \ref{appendix:regionsdefined} Table \ref{tab:region_definitions}), for the purposes of mass calculations presented in Table \ref{tab:mass}.}}
\label{fig:C34S}
\end{figure*}

A wealth of archival molecular line data is available for HH212, among which \reply{SO (9$_8$-8$_7$), SO$_2$ (8$_{2,6}$-7$_{1,7}$), and C$^{34}$S (7-6) are known wind and cavity tracers} \citep{Taboneetal2017_SO_SO2,Codellaetal2018_HDO} (see \S \ref{sub:moleculartracers} and Appendix \ref{appendix:othermolecules} for further discussion). The molecular line data, specifically at comparable angular resolution, provides important kinematic information and can be used to better understand the nature of the extended dust emission. To this end, Figure \ref{fig:C34S} presents the re-imaged Band 9 data overlaid on the SO and C$^{34}$S data from \citet{Taboneetal2017_SO_SO2}. We note that the clearly delineated SO wind and C$^{34}$S cavity walls (denoted by the dashed green contour) coincide with the dust extensions seen in Bands 9 (and 7); we point the reader to Appendix \ref{appendix:othermolecules} for complementary figures that show molecular lines in this region for context. Towards the south-southwest, the coincidence between continuum and especially SO is excellent. Meanwhile, the branch of the extended emission towards the southeast of the disk has a component which aligns with the disk major axis and is clearly beyond the outflow cavity \replyfive{(likely envelope, see Appendix \ref{appendix:othermolecules} for C$^{17}$O  map)}, but also contains a mostly southern component which appears to be inside the outflow cavity and not unambiguously coincident with envelope alone. 

To test whether the observed extensions are oriented in a preferred direction we carried out a quantitative analysis (Appendix D). For Band 7, the most significant residual sector is oriented from 81° to 171° east of north ($p = 0.0014$). For Band 9, the corresponding sector spans from 63° to 153° east of north ($p = 0.0006$). Hence, the analysis in both bands identifies a statistically significant residual excess oriented towards the south-east, indicating that this preferred direction is unlikely to result from random fluctuations.

\subsection{The spectral index}
\label{sub:spectralindex}
\begin{figure}[ht]
\centering
\includegraphics[trim=0 0 120 0,clip,width=\columnwidth]{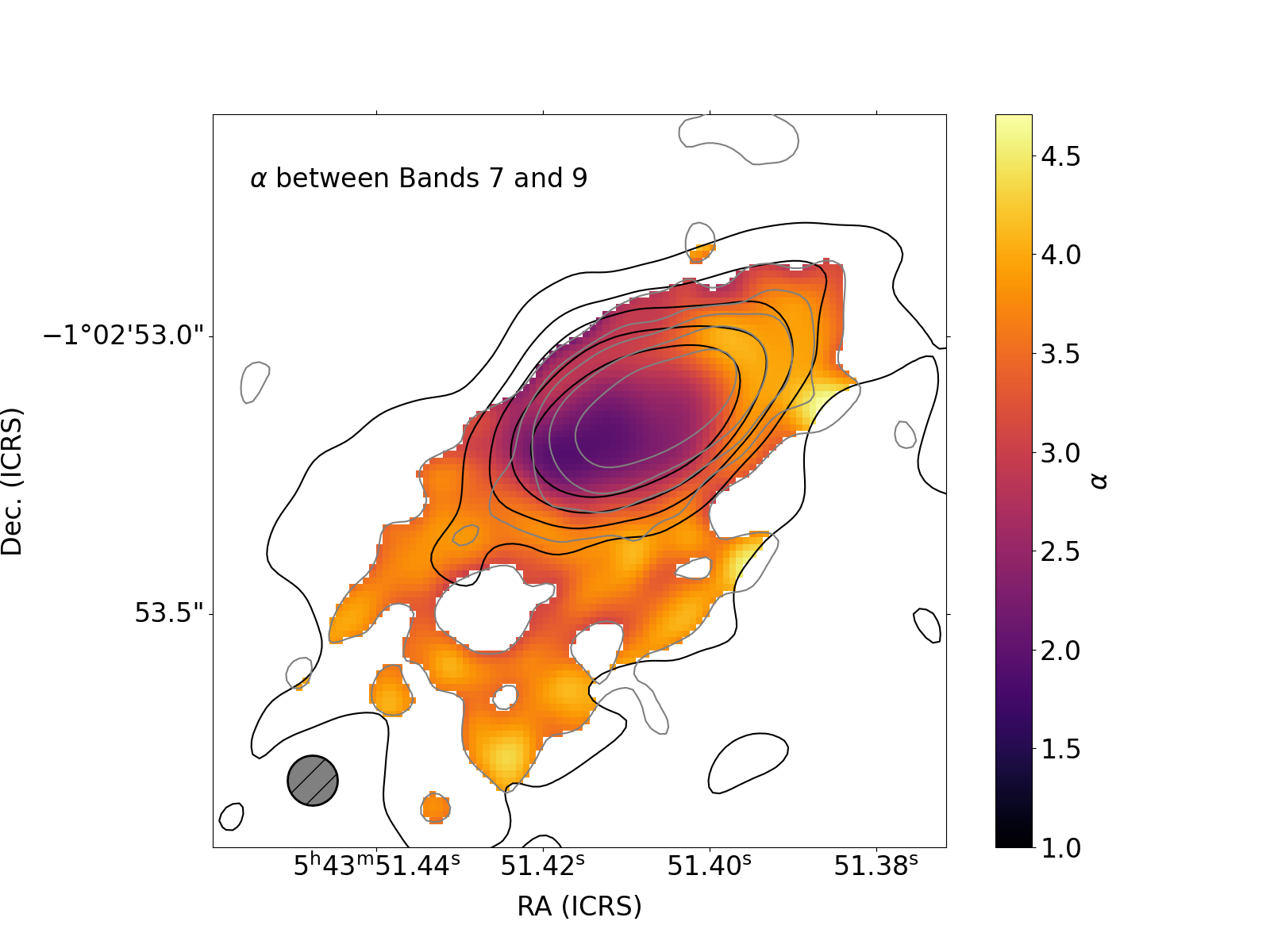}
\caption{The spectral index $\alpha$ between ALMA Bands 7 and Band 9 (color). Both data sets have been smoothed to a 0.09$\arcsec$\ beam size (shown as hatched circle in lower left). Overlaid are contours for the Band 7 (black) and Band 9 (grey) continuum emission. The contours are drawn at 4$\sigma$, 8$\sigma$, 12$\sigma$, 20$\sigma$ and 40$\sigma$. 4$\sigma$ in Band 7 corresponds to 0.56 mJy/beam and to 8.8 mJy/beam in Band 9 for a 0.09$\arcsec$\ beam.}
\label{fig:spectralindex}
\end{figure}
\begin{figure*}[ht]
\centering
\includegraphics[trim=120 0 120 0,clip,width=0.49\textwidth]{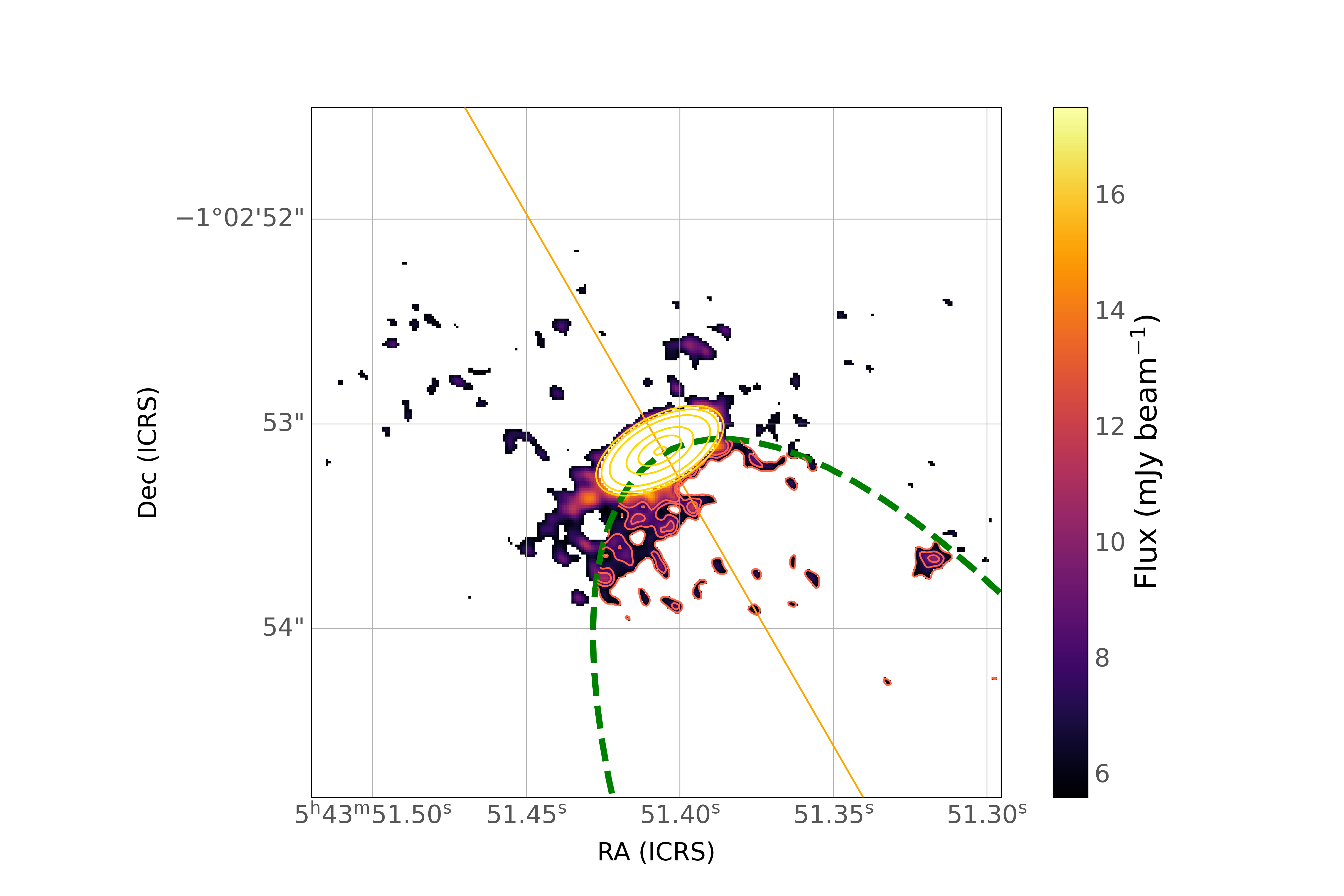}
\includegraphics[trim=120 0 120 0,clip,width=0.49\textwidth]{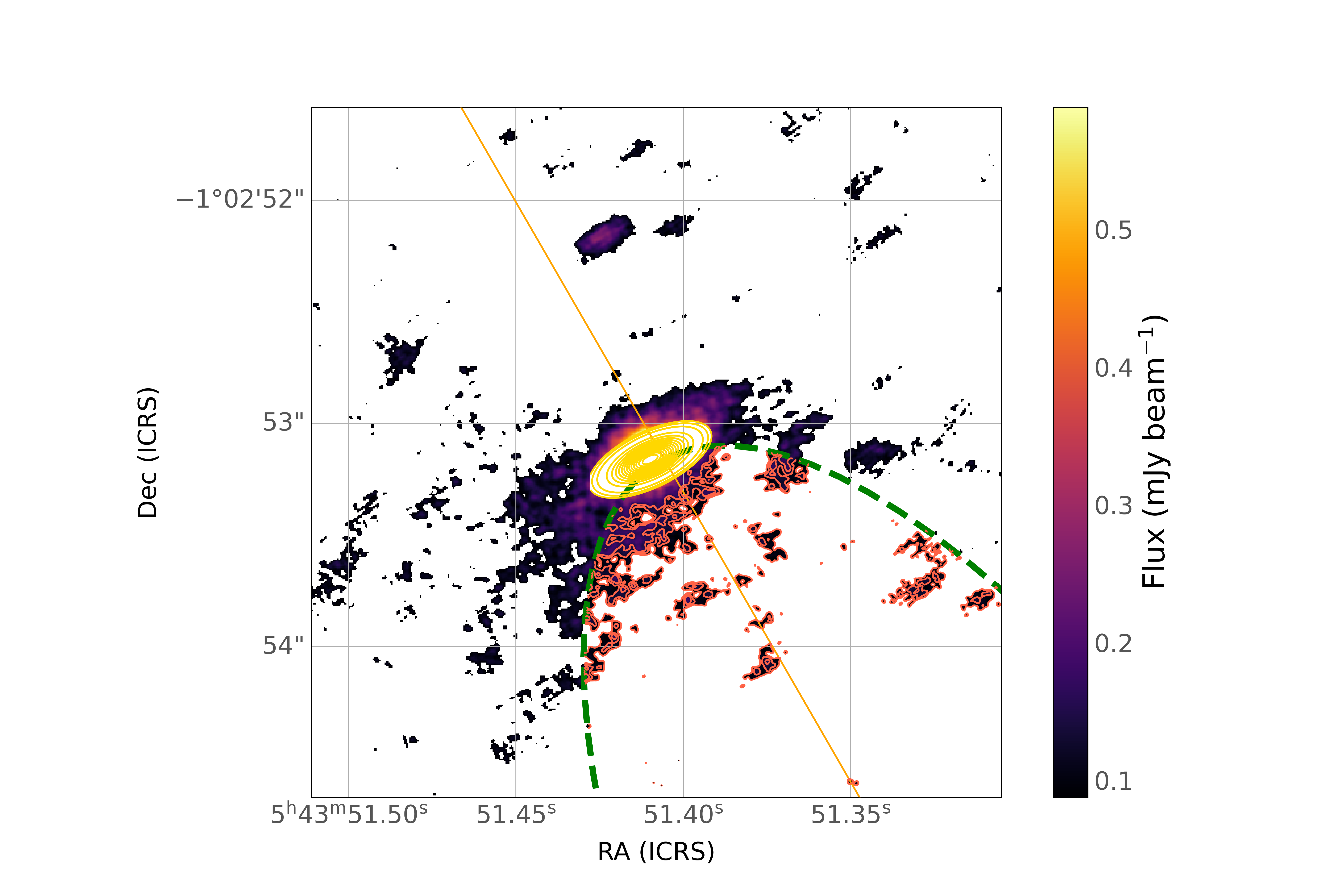}
\caption{Band 9 (\emph{left}) and 7 (\emph{right}) continuum maps (color) at $\geq 3 \sigma$ and masked to remove the disk component. A two-dimensional Gaussian was fit to the central emission (yellow contours, at levels of $3\sigma$, $5 \sigma$, $10 \sigma$, and then intervals of $20 \sigma$) and subsequently subtracted from the images.  Orange-red contours show $3, 4, 5\sigma$ levels. As before, the dashed green contour encloses the region where significant C$^{34}$S emission is observed to the southwest, and the orange line denotes the bisection of C$^{34}$S emission lobes.}
\label{fig:contmasking}
\end{figure*}

To characterize the dust emission and its extension, we calculate the observed spectral index from the dust thermal emission. The dust spectral index, commonly denoted as $\alpha$, describes the dependence of the continuum flux on frequency or wavelength. Under the assumption that the continuum flux is of the form $S_\nu \propto \nu^\alpha$, we calculate the spectral index $\alpha$ according to \citep[e.g.][]{Beckwithetal1990}:
\begin{equation}
  \alpha = \frac {\log_{10} ({S_{\nu_1}/S_{\nu_2}})}{\log_{10} ({\nu_1/\nu_2})},
  \label{eq:alphamm}
\end{equation}
where $S_{\nu_1}$ is the flux at frequency ${\nu_1}$ (the lower frequency band), and $S_{\nu_2}$ is the flux at the frequency ${\nu_2}$ (the higher frequency band). In the optically thin limit, the flux can be related to the physical properties of the emitting dust via
\begin{equation}
  S_\nu = \frac{\kappa(\nu) M_{\rm dust} B_{\nu}(T)}{d^2},
  \label{eq:opticallythinflux}
\end{equation}
where $d$\ is the distance to the source, $\kappa(\nu)$ is the frequency-dependent dust opacity, $B_\nu(T)$ is the Planck function, $M_{\rm dust}$ is the mass of the emitting dust, and $T$ is the temperature of the emitting dust \citep[e.g.][]{Draine2006}. If the opacity can be represented by a power-law in frequency, $\kappa(\nu) \propto \nu^\beta$, which is reasonable at millimeter and sub-millimeter wavelengths under the Rayleigh-Jeans approximation, then the spectral index is related to the opacity through $\alpha = 2 + \beta$, and $\beta$ depends on the maximum dust grain size, composition, and the size distribution \citep[e.g.][]{Testietal2014_PPVI}.

Given the availability of multi-frequency data for HH212, we can calculate the spectral index as a function of position using the flux values from each image \citep[see also][]{Linetal2021}. In particular, we are interested in the spectral index between Bands 7 and 9, which show extensions in the dust emission beyond the known disk of HH212.  Figure \ref{fig:spectralindex} presents a map of the spectral index, where Eq.\ (\ref{eq:alphamm}) was applied to each pixel in which both Bands 7 and 9 continuum emission exceeds 4$\sigma$. The spectral index is notably low (minimum value of $\sim$1.9) in the disk region, and relatively high in the regions of the extensions (maximum value of $\sim$4.8). 
In order to investigate the mean spectral index in different parts of the disk-envelope-outflow structure, we divide the map into three ``regions,'' shown visually in Figure \ref{fig:C34S}, and explained in more detail in Appendix \ref{appendix:regionsdefined}.  This also allows us to distinguish areas where the assumption of optically thin dust continuum emission is more likely to hold.

Considering emission greater than 4$\times$ the rms noise level, we find the mean $\alpha$ values for these three regions to be 3.38 $\pm$ 0.57 (Region 1), 3.63 $\pm$ 0.35 (Region 2), and 3.74 $\pm$ 0.29 (Region 3), respectively, where the uncertainties on the means are given by the standard deviations of the results.  Given that the HH212 disk is optically thick at Band 7 and 9, the spectral index results for Region (1) are unreliable (see Appendix \ref{appendix:regionsdefined}), and we choose to focus on the average $\alpha$ values from Regions (2) and (3). Using $\alpha = 2 + \beta$, we obtain opacity power-law indices for Regions (2) and (3) of $\beta = 1.63 \pm 0.35$ and $1.74 \pm 0.29$, respectively. 

For comparison with other studies on the same source, \replyfour{\cite{Leeetal2008} used Submillimeter Array observations and derived a value of $\beta \simeq 0.6$ on scales of $\sim$0.3\arcsec, which includes the HH212 disk contribution. Recently, using ALMA and VLA data, \citet{Hu2025} made spectral index maps of the disk region of HH212, and, based on multi-band disk modeling with scattering, they found $\beta \simeq 1.2$ between ALMA Band 7 and VLA Q-Band.  They excluded the Band 9 data from the determination of the opacity
spectral index $\beta$ due to its relatively lower angular resolution, as well as higher contamination from the envelope versus disk.  On the other hand, \citet{Hu2025} measure the spectral index $\alpha=2.42$ in the HH212 disk using ALMA Bands 3-9, which seems to indicate that the disk is optically thick.}  Meanwhile, the value of $\beta$ in the diffuse interstellar medium (ISM) is $\simeq 1.7-1.8$ \citep[e.g.][]{Riccietal2010,Juvela2015}, although some variations may arise due to dust composition, porosity, geometry and size distribution \citep{Testietal2014_PPVI}. 

\subsection{Estimating the Dust Mass}
\label{sub:dustmass}

To investigate how much dust there is outside of the HH212 disk, and whether it is significant relative to the mass of the disk, we estimate the amount of mass in the regions described above by Regions (2) and (3) \replyfour{in the Bands 7 and 9 images}. We adopt the same region masks as before, but we consider continuum emission $> 3\sigma$, $> 4\sigma$ and $> 5\sigma$ threshold levels for a more complete analysis (see ``Sigma Cut'' in Table \ref{tab:mass}). We direct the reader to Figure \ref{fig:contmasking} for a visual representation of the emission structure in both bands.

The mass of the emitting dust is calculated according to
\begin{equation}
M_\lambda= \frac{Scd^2}{B_\lambda(T)\kappa_{\lambda}\lambda^2},
\label{eq:Mla}
\end{equation}
where $S$ is the flux (in mJy), $d$\ is the distance to the source (in parsecs), $\kappa_\lambda$\ is the wavelength dependent dust opacity (cm$^2$ g$^{-1}$), $c$ is the speed of light, and $ B_\lambda(T)$\ is the Planck function. Equation \ref{eq:Mla} differs from, e.g., the form found in \citet{Motte01,Bjerkeli19} due to the use of the wavelength-based form of the Planck function (i.e.\ different by a factor of c/$\lambda^{2}$). We note that Eq.\ \ref{eq:Mla} is only valid in the optically-thin regime.

To proceed, we adopt a (constant) temperature of $T = 59$ K outside of the disk \cite[based on][]{Linetal2021}. We expect the temperature beyond the HH212 disk to be in the range of \replyfour{50 K - 140 K} \citep{Linetal2021}, but to be on the lower end in the extended regions we are focusing on. We appreciate that the mass we estimate depends on the temperature through the Planck function; a factor of three difference in temperature changes the value of the Planck function (and therefore our mass estimates) by a factor of $\sim$3 at sub-millimeter wavelengths.

Arguably, the dust opacity is a much larger source of uncertainty in the mass calculation than the temperature; the exact composition, grain size distribution, and even shape of the dust grains influence the opacity. Fortunately, in the sub-millimeter and millimeter wavelength regime, dust opacities are well-fitted by a power-law, $\kappa_\lambda \sim \kappa_{\lambda,0} (\lambda / \lambda_0)^{-\beta}$. With values of $\beta$ in hand (Sect.\ \ref{sub:spectralindex}), the next thing we need is a value for the pre-factor $\kappa_{\lambda, 0}$.

However, as we do not know the true properties of the dust in HH212, we adopt two well-known opacity examples from the literature. First, we consider the opacity of a bare, compact, amorphous silicate/graphite dust mixture \replytwo{\citep{Draine2003a,Draine2003b} with grain sizes between 0.005 -- 0.25 $\mu m$ and size distribution $n(a) \propto a^{-q}$, with $q=3.5$ \cite[similar to][]{MRN1977}. It provides a dust opacity of $\kappa_{\lambda, 0}$ at 1 mm of $0.35$ cm$^2$ g$^{-1}$. The particular value of $0.35$ cm$^2$ g$^{-1}$ at 1 mm was determined using the dust properties just described as input to the \texttt{optool} software package \citep{optool2021ascl}}\footnote{\url{https://github.com/cdominik/optool}}. Second, we consider the opacity of \citet[][hereafter O\&H94; \replytwo{their Table 1}]{OssenkopfHenning1994} for \replytwo{porous grains with thin ice mantles}, which gives \replytwo{a dust opacity} $\kappa_{\lambda, 0}$ at 1 mm of $\simeq$ \replytwo{0.78} cm$^2$ g$^{-1}$. This opacity is often used for protostellar cores, and includes ice mantles with volumes set to half of the refractory component.

\begin{table*}[ht]
\setlength{\tabcolsep}{3pt}
\footnotesize
  \begin{center}
    \caption{Estimates of the dust mass in the extended dust emission.}
    \label{tab:mass}
    \begin{tabular}{ccccccccc}
    \hline
\hline
\textbf{Band} & \textbf{Region} & \textbf{$\sigma$ Cut} & \textbf{Flux} & \textbf{RMS} & $\mathbf{M}$ (O\&H94) & $\mathbf{M}$ (\replytwo{D03}) & $\frac{\mathbf{M}}{\mathbf{M_{\rm d}}}$ (O\&H94) & $\frac{\mathbf{M}}{\mathbf{M_{\rm d}}}$ (\replytwo{D03}) \\
 & & & mJy & mJy & $M_{\rm Jup}$ & $M_{\rm Jup}$ & \% & \% \\
\hline
9 & outside disk (Reg.~2) & 3+ & 1848 &    1.86 &   75.08 $\pm$ 21.99 &  \replytwo{166.59 $\pm$ 48.80} &  143.34 $\pm$ 41.99 &  \replytwo{318.04 $\pm$ 93.17} \\
\reply{9} & \reply{outside disk} (Reg.~2) & \reply{4+} &  \reply{467.09} &    \reply{1.86} &   \reply{18.98 $\pm$ 5.56} &   \replytwo{42.11 $\pm$ 12.34} &   \reply{36.24 $\pm$ 10.62} &   \replytwo{80.40 $\pm$ 23.56} \\
9 & outside disk (Reg.~2) & 5+ &  181.13 &    1.86 &    7.36 $\pm$ 2.16 &   \replytwo{16.33 $\pm$ 4.79} &   14.05 $\pm$ 4.12 &   \replytwo{31.18 $\pm$ 9.14} \\
9 & inside cavity (Reg.~3) & 3+ &  287.79 &    1.86 &   10.67 $\pm$ 2.59 &   \replytwo{23.67 $\pm$ 5.75} &   20.36 $\pm$ 4.94 &   \replytwo{45.18 $\pm$ 10.97} \\
\reply{9} & \reply{inside cavity} (Reg.~3) & \reply{4+} &  \reply{138.56} &    \reply{1.86} &    \reply{5.14 $\pm$ 1.25} &   \replytwo{11.39 $\pm$ 2.77} &    \reply{9.80 $\pm$ 2.38} &   \replytwo{21.75 $\pm$ 5.29} \\
9 & inside cavity (Reg.~3) & 5+ &   74.45 &    1.86 &    2.76 $\pm$ 0.67 &    \replytwo{6.12 $\pm$ 1.49} &    5.27 $\pm$ 1.29 &   \replytwo{11.69 $\pm$ 2.85} \\
7 & outside disk (Reg.~2) & 3+ &  116.73 &    0.03 &   29.69 $\pm$ 3.09 &   \replytwo{65.87 $\pm$ 6.85} &   56.67 $\pm$ 5.89 &  \replytwo{125.75 $\pm$ 13.07} \\
\reply{7} & \reply{outside disk} (Reg.~2) & \reply{4+} &   \reply{63.36} &    \reply{0.03} &   \reply{16.11 $\pm$ 1.68} &   \replytwo{35.75 $\pm$ 3.72} &   \reply{30.76 $\pm$ 3.20} &   \replytwo{68.26 $\pm$ 7.10} \\
7 & outside disk (Reg.~2) & 5+ &   39.86 &    0.03 &   10.14 $\pm$ 1.05 &   \replytwo{22.49 $\pm$ 2.34} &   19.35 $\pm$ 2.01 &   \replytwo{42.94 $\pm$ 4.46} \\
7 & inside cavity (Reg.~3) & 3+ &   28.89 &    0.03 &    7.11 $\pm$ 0.61 &   \replytwo{15.78 $\pm$ 1.36} &   13.57 $\pm$ 1.17 &   \replytwo{30.12 $\pm$ 2.59} \\
\reply{7} & \reply{inside cavity} (Reg.~3) & \reply{4+} &   \reply{16.55} &    \reply{0.03} &    \reply{4.07 $\pm$ 0.35} &    \replytwo{9.04 $\pm$ 0.78} &    \reply{7.78 $\pm$ 0.67} &   \replytwo{17.26 $\pm$ 1.49} \\
7 & inside cavity (Reg.~3) & 5+ &    9.59 &    0.03 &    2.36 $\pm$ 0.20 &    \replytwo{5.24 $\pm$ 0.45} &    4.51 $\pm$ 0.39 &   \replytwo{10.00 $\pm$ 0.86} \\
\hline
\end{tabular}
  \end{center}
    \tablecomments{Mass estimated from the continuum emission in each of the defined regions (Sect.\ \ref{sub:spectralindex}). Mass estimates used are calculated from the combination of the $\beta$ values calculated in Sect.\ \ref{fig:spectralindex} with the scaling constant given by either \replytwo{\cite{Draine2003a,Draine2003b}} or \cite{OssenkopfHenning1994} opacities at 1 mm. The $\beta$ value for Region (2) is $1.63 \pm 0.35$, and $1.74 \pm 0.29$ for Region (3). A disk mass of $0.05M_\odot$ was used. Uncertainties in the mass are given by the standard method of propagating uncertainties in the flux (i.e.\ RMS) and $\beta$ values. \replyfour{Figure \ref{fig:mass_histogram} shows the corresponding masses in histogram form.}}
\end{table*}

\begin{figure*}[ht]
    \centering
    \includegraphics[width=0.48\textwidth]{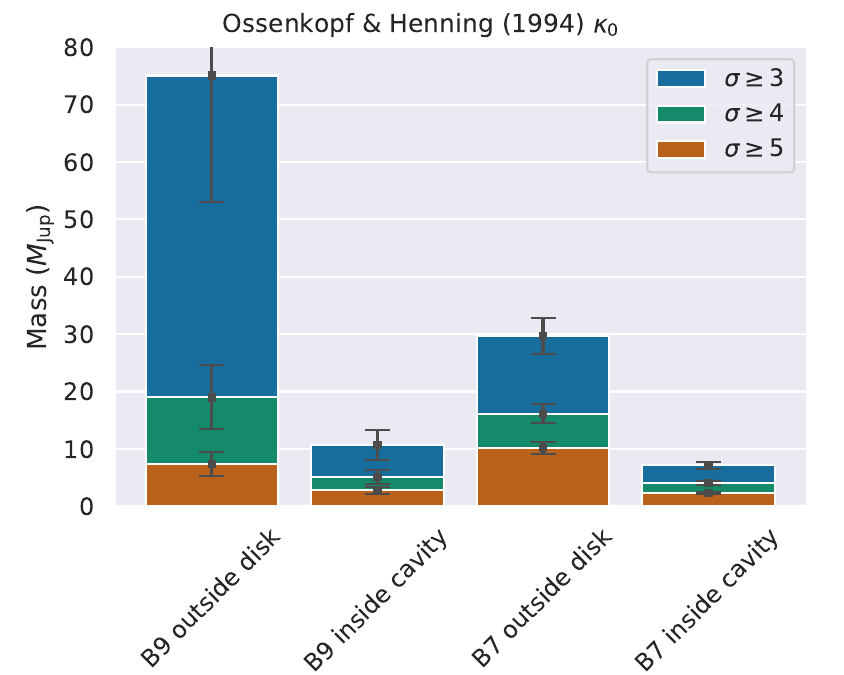}\includegraphics[width=0.48\textwidth]{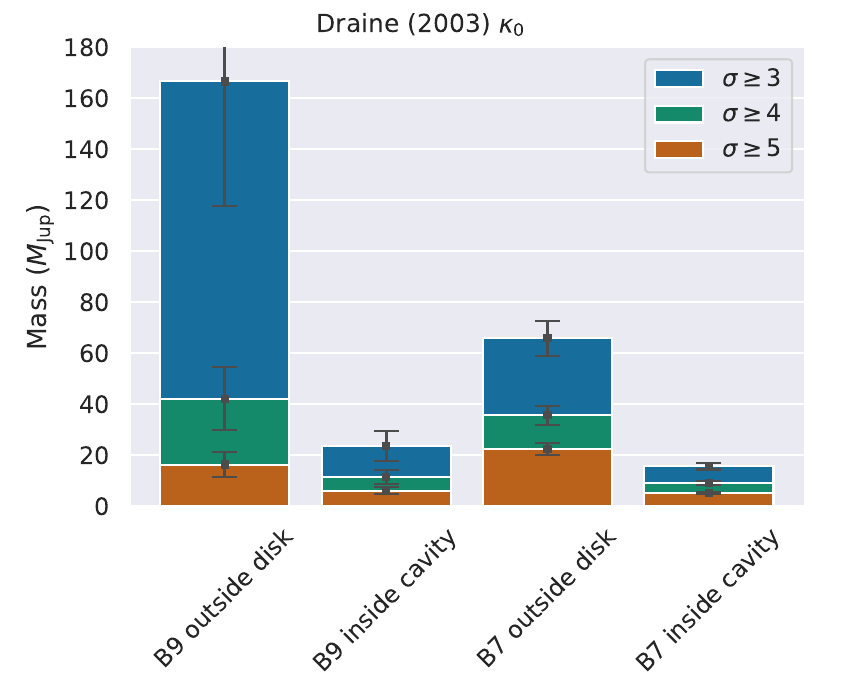}
    \caption{Histograms of the estimated masses \replyfour{of indicated regions observed with Bands 7 and 9}, using Eq.\ \ref{eq:Mla} and the data in Table \ref{tab:mass}. Uncertainties in mass are calculated using the RMS values ($\Delta M \propto$ RMS) \reply{and uncertainties in $\beta$ values,} and shown as black \reply{error bars}. \textit{Left:} Estimated mass adopting $\kappa_\lambda \propto \kappa_{\lambda, 0} (\lambda / 1 \mathrm{mm})^{-\beta}$, with $\kappa_{\lambda, 0}$ = 0.078 cm$^2$ g$^{-1}$ \citep{OssenkopfHenning1994}. \textit{Right:} Estimated mass adopting \replytwo{ $\kappa_{\lambda, 0}$ = 0.35 cm$^2$ g$^{-1}$ \citep{Draine2003a,Draine2003b}}.  Note the difference in the scaling of the y-axis between panels.}
    \label{fig:mass_histogram}
\end{figure*}

We employ the two different values of $\kappa_{\lambda,0}$ for \reply{Regions} (2) and (3) with the $\beta$ value estimated for that region to calculate the mass from Eq.\ \ref{eq:Mla}. The results of the calculations are found in Table \ref{tab:mass}\replyfour{, and presented in histograms in Fig. \ref{fig:mass_histogram}}. We adopt a distance of 400 pc for HH212 \citep{Kounkel17}.  \replyfour{As mentioned in \S \ref{sub:emission}, the images in different bands do not all have the same sensitivity in terms of dust mass; specifically, using the same temperature and opacity assumptions, the smoothed Band 7 and 9 data are comparable within a factor of $\sim1.5$, whereas the Band 9 data is more sensitive to dust mass than the Band 3 by a factor of $\sim6-7$, depending on the chosen $\beta$ value.}

Several different values for the mass of the disk in HH212 can be found in the literature. The values vary depending on whether the optically thin assumption is made (i.e.\ Eq.\ \ref{eq:Mla}), and the wavelength of the observations (e.g.\ 0.87 mm versus 9 mm; \citealt{Tobinetal2020}). Here, we adopt a total disk mass of $0.05M_\odot$ \citep{Lee2017SciAdv}, even though this may be an upper limit, particularly considering that the protostellar mass is $\simeq0.25M_\odot$ \citep{Lee2017ApJ}. Table \ref{tab:mass} also presents the amount of mass in dust in each region expressed as a percentage of the adopted total disk mass.


\section{Interpretation \& Discussion}
\label{sec:interpretation}

\subsection{Dust morphology}
\label{sub:dustmorph}

Thermal dust emission from a protoplanetary accretion disk (without substructure) is expected to have the shape of a two-dimensional Gaussian when projected on to the sky with some inclination angle. In some cases, disks can appear to have a ``boxy'' shape -- e.g.\ as presented by \citet{Linetal2021} for HH212 in Bands 3, 6, and 7 -- rather than a Gaussian shape.  This is, however, an idealized picture. Recent observations, in particular with ALMA, have demonstrated that dust substructures (gaps and rings) within disks \citep[e.g.][]{Andrewsetal2018_DSHARP1} are common, although the overall structure can still be fit, to first-order at least, by a two-dimensional Gaussian in most cases. Deviations from ``Gaussianity'' in disks have also been observed in sources where there is a clear interaction with an envelope, and material is actively accreting onto the disk from the surrounding nascent material \citep[][and references therein]{Pinedaetal2022_PPVII}.  \replytwo{The observations of several class 0 protostellar cores by \citet{Legouellec2020} demonstrated that it is not uncommon for dust emission to coincide with either cavity walls and/or envelope structures, and hence reveal asymmetric structures in young sources when observed with sufficient resolution and sensitivity.}

While continuum emission itself does not provide kinematic information, molecular line emission coincident with dust continuum structures can serve as a proxy to constrain whether material is falling inward or being ejected outward, in the current context, either towards the disk or away as part of an outflow.  As shown in \S \ref{sub:emission}, and especially in Figures \ref{fig:C34S} and \ref{fig:contmasking}, the dust continuum emission for HH212 deviates from Gaussianity at levels of $3-10\sigma$ in \replyfive{Bands} 7 and 9.  Notably, some of this non-Gaussian emission extends into and coincides with the southern outflow launching region, as seen in SO and C$^{34}$S (Figure \ref{fig:C34S}; see also below). Based on our estimates of the mass in this region (for emission with $> 3\sigma$) in Sect.\ \ref{sub:dustmass} (``Region 3''; Table \ref{tab:mass} and Fig.\ \ref{fig:mass_histogram}), we interpret this as potential evidence that a small but significant amount of dust (relative to the mass of Jupiter) is being carried away by the outflow. 

To test whether the deviations could be due to the ``boxy'' structure suggested by \citet{Linetal2021}, we subtracted their model fit (Z.-Y.\ D.\ Lin; priv. communication; see their Fig.\ 10) from our data, but the extensions that we report here persist beyond any reasonable contribution from the spatial extent of the disk structure. \replythree{A quantitative analysis of the residuals further suggests that the excess emission is preferentially oriented towards the south-east (Appendix D), strengthening the case that it traces non-disk material.} \replytwo{We emphasize that the portion of dust that could be carried away by the outflow corresponds to the region outlined by and within the cavity structure, coincident with SO, and within the C$^{34}$S boundary.  Whereas other sources may have revealed dust coincident with cavity walls \citep{Legouellec2020}, the correspondence of dust within cavity structures has not been previously established.} 

We note that there is also extended dust emission to the east of the source, along the disk major axis, and that there is also a small but significant amount of mass present there (``Region 2''; Table \ref{tab:mass} and Fig.\ \ref{fig:mass_histogram}). However, this emission lies outside the known outflow cavity structure, and is likely not related to the outflow itself. Instead, it is plausible that the extended emission along the disk major axis may be remnants of an envelope accretion structure, where material is accreting onto the outer edges of the disk.  Accordingly, in order to distinguish dust specifically in the wind region from that exclusively in the envelope, we define ``\reply{Region} 3'' based on a previously-known wind tracer (note that ``\reply{Region} 2'' encompasses ``\reply{Region} 3'').

\subsection{Constraints provided by molecular line emission}
\label{sub:moleculartracers}

The jet from HH212 was first discovered by \cite{Zinneckeretal1998} via H$_2$ line emission in the near-IR at 2.12 $\mu$m. Subsequent observations demonstrated that the jet is episodic, and moving at $\simeq$170 km s$^{-1}$ \citep[][and references therein]{reipurthetal2019}. Through observations of SiO \citep{Lee2017NatAs,Tabone20,LeeLietal2022}, the base of the highly collimated jet is known to coincide with the position of the protostar and the innermost regions of the HH~212 disk. Line observations of other molecular species, however, provide us with the most relevant information about the properties of the slower, wide-angle wind.

For example, emission from $^{12}$CO (J=2-1) traces the outflow cavity walls and entrained material, particularly within $\sim 5 \arcsec$ of the protostar \citep{Leeetal2006,Cabritetal2012}, and coincides with the region where we observe extended dust emission to the south of the disk. While $^{12}$CO (J=2-1) emission has been used to constrain the outflow launching region in other sources at very high angular resolution \citep[e.g.][]{bjerkeli2016,devalonetal2020_dgtaub}, only lower resolution observations are currently available for this particular transition in HH212 ($\gtrsim 0.44\arcsec$; \citealt{Leeetal2006, Cabritetal2012}). The $^{12}$CO (J=3-2) emission towards HH212 has, meanwhile, been observed \citep{Leeetal2015} at resolutions on the order of 0.5$\arcsec$. At high velocities ($|v - v_{\rm sys}| \gtrsim 8$ km s$^{-1}$), the J=3-2 emission towards HH 212 traces the same component of the jet as H$_2$ and SiO. At low velocities close to the systemic velocity, the J=3-2 transition instead traces the outflow cavity walls, similar to the J=2-1 transition. 

Looking instead at the emission from SO and SO$_2$, \citet{Podio2015_SO,Taboneetal2017_SO_SO2,Lee2018b} have found that these molecules trace not only the outflow cavity walls, but also show evidence for rotation in the outflow itself. We suggest that the \reply{SO and SO$_2$ represent the material launched from a disk wind within a cavity outlined by} C$^{34}$S \citep{Codellaetal2014_HH212,Taboneetal2017_SO_SO2,Tabone20,LeeTaboneetal2021}. As explained in \S \ref{sub:spectralindex}, we have adopted the edge of the \reply{C$^{34}$S and SO} emission as the marker of the boundary between outflowing material and other kinematic components in the system (Figs.\ \ref{fig:C34S} and \ref{fig:appendix_soso2}). The dust extensions we see in Band 9 (``Region 3'') are spatially coincident \reply{particularly with the SO emission}, suggesting that the dust extension we observe is potentially dust particles being carried away by the disk wind. Given the evidence, we think it is unlikely that the dust emission coincident with, and inside, the cavity walls can be related to infall, but is instead being entrained or pushed away from the disk and protostar. 
\reply{Kinematic evidence of the dust itself would be required to confirm or refute this claim.} 

As mentioned before, we also observe continuum emission outside the disk, roughly along the disk major axis (included in the ``\reply{Region} 2'' region), but this emission is not coincident with any of the molecules that trace the jet, wind or cavity walls. It is, however, coincident with low-velocity C$^{17}$O and HCO$^+$ emission \citep{Codellaetal2014_HH212, Lee2017ApJ,LeeTaboneetal2021}, and we reiterate that it could therefore be infalling envelope material in that particular region. \reply{ALMA archival image products are available for the optically thin isotopologue C$^{17}$O (J=3-2) towards HH 212 (see Figure \ref{fig:appendix_soso2}), allowing us to mark the extent of the envelope emission, as well as its substructures that trace the cavity walls at high angular resolution ($\sim0.1\arcsec$).}
Together with emission from HCO$^+$ ($J=4-3$), it is clear that HH212 maintains an envelope with a flattened, rotating structure that extends all the way down to the nominal disk \citep[see also][]{Codellaetal2014_HH212,Lee2017ApJ}. At velocities $1.5 \leq |v - v_{\rm sys}| \leq 3.0$ km s$^{-1}$, the HCO$^+$ ($J=4-3$) emission appears to trace both the cavity walls and material \emph{outside} the cavity walls (as, e.g., delineated by low velocity $^{12}$CO J=3-2). \reply{C$^{17}$O shows such a structure, especially at lower velocities within about 1 km s$^{-1}$ of $v_{\rm sys}$ (see Figure \ref{fig:appendix_soso2}).}  \replyfour{On} the other hand, several complex organic molecules (COMs) have also been observed towards HH212, with most emission appearing immediately above the continuum disk, seemingly tracing a rotating ``atmosphere'' above the disk \citep{Lee2017ApJ,Codellaetal2016_HDO,Codellaetal2018_HDO}.

\subsection{Dust in the wind?}
\label{sub:dustinthewindornot}

What is the origin of the extended dust emission that we observe coincident with \replytwo{and within} the outflow cavity walls? The higher flux in Band 9 in this region relative to the other ALMA bands (Fig.\ \ref{fig:allbands_cont}) results in a spectral index in this region (Fig.\ \ref{fig:spectralindex}) that is significantly higher than in the disk (cf.\ $\alpha \simeq 3$; \citealt{Linetal2021}). This suggests the presence of smaller dust grains ($a < 0.1$ mm; see Sect.\ \ref{sub:spectralindex}), under the assumption that the dust opacity $\kappa(\nu) \propto \nu^{\beta}$, where the spectral index is $\alpha = 2 + \beta$. Indeed, we estimate a value of $\beta = 1.74 \pm 0.29$ in the dust extension, consistent with the value of $\beta$ for the diffuse ISM, to within the uncertainties. \replyfour{As mentioned in \S \ref{sub:spectralindex}, the value we measured outside the disk is somewhat higher than the disk value of $\beta=1.2$ determined by \citet{Hu2025}.  Higher $\beta$ values indicate that the extended emission traces dust with different effective properties than the scattering-dominated, optically thick disk; however, a direct quantitative comparison is limited by different methodologies.}  In studies of $\beta$\ in other protostellar envelopes, \citet{Galametzetal2019,miotello2014} have measured values of $\beta<1$ for some sources in their samples, which they refer to as ``surprisingly low.'' However, \citet{Galametzetal2019} report radial variations in the envelopes of several objects they studied, and at larger radii they present values of $\beta$\ consistent with our findings; \citet{bracco2017} also found increasing $\beta$ profiles at larger radii for the two protostellar cores in their sample.  We therefore suggest that the spatial assessment (rather than average) of spectral indices is imperative for a better understanding.

 \reply{The findings of larger $\beta$\ values in the path of the wind} can be interpreted in two ways. On the one hand, it is expected that small grains will be preferentially lifted into the wind due to their smaller inertia (relative to larger grains), and the measured value of $\beta$ in the extension supports this. \replytwo{In contrast, given that our $\beta$ value is consistent with the accepted value for the diffuse ISM to within our uncertainties, it suggests we might instead be observing emission from unprocessed ISM dust.} Although dust can plausibly grow within protostellar envelopes, causing $\beta$ to decrease, a measurement of the spectral index in the envelope \replytwo{would be} needed to confirm this. To the best of our knowledge, there is no measured value in the literature for $\alpha$ or $\beta$ purely for the envelope of HH212; \reply{on the other hand, it seems observationally impossible to distinguish dust that has exclusively resided in the envelope from dust that has been carried (via a wind) from the disk, through the envelope in order to test the hypothesis}.  \replythree{A very interesting recent study by \citet{Sabatini2025} does show one case, the Class I source L1551 IRS5, in which large grains are found in the direction of one cavity wall.  A similarity with that study is that in both the cases of HH212 and L1551 IRS5, the dust seems to trace the shape of the cavity.  In contrast, an important finding of \citet{Sabatini2025} is signs of grain growth.}

\reply{In the context of, e.g., \cite{Riccietal2010,Testietal2014_PPVI}, based on the estimated value of $\beta$, it is apparent that the dust we are observing outside of the HH212 disk has not significantly grown (in contrast to the dust within the HH212 disk which has \replyfour{$\beta \simeq 1-1.2$; \citealt{Linetal2021,Hu2025})}, \replyfour{and we estimate dust grain sizes} $a<0.1$ mm (see Fig.\ 3 of \citealt{Riccietal2010} or Fig.\ 4 of \citealt{Testietal2014_PPVI}).  \replyfour{The dust grain sizes may be even smaller, given that \citet{Hu2025} show that $a<0.13$ mm should be an upper limit for the maximum grain size in the HH212 disk, which does show signs of grain growth.}  We note, also, that the maximum measured value in our map is $\beta \simeq 2.8$ ($\alpha \simeq 4.8$; see Fig.\ \ref{fig:spectralindex}). Although such a high $\beta$ is very localized to one region in the west direction, comparing it to \cite{Riccietal2010}, the maximum grain size in this small region can be constrained to 0.1 -- 0.5 mm. Theoretical models, meanwhile, find that winds can likely only launch small grains ($\lesssim 10 \mu\mathrm{m}$) from the disk into the wind \citep{Franzetal2020_dustphotoevapI,Franzetal2022a_imaging,BoothClarke2021_dusttransport, Giacaloneetal2019_dusttransport}, which is consistent with our spectral index (hence $\beta$) measurements for HH212.} \replythree{Comparing with the Class I case of L1551 IRS5, then it is plausible that while grain growth has not yet begun at the evolutionary stage of HH212, it may occur more significantly during the transition from Class 0 to I.}

Dust continuum observations do not contain kinematic information that could be used to directly constrain whether the emission we observe is a part of the outflow. Instead, we could compare with emission from gas that is released from the dust as a result of heating or shocks. Examples of these tracers could be SiO \citep[e.g.][]{Lee2017NatAs} or CH$_3$OH \citep[e.g.][]{Bianchietal2017_HH212methanol}, but neither of those observations show emission coinciding with the dust extension \replyfour{that we present in this work.  A} more direct means to study the dust kinematics would involve proper motion measurements of dust emission over a period of several years. This would enable us to compare with kinematics of the outflow, specifically showing whether dust coincident with the wind is moving away from the disk or infalling towards the disk, or instead moving with the envelope. \replytwo{In contrast, this could instead help constrain whether the dust we see pertains instead to projection effects of a slightly asymmetric envelope}. While such a dust proper motion study will be challenging, particularly at higher frequencies like \replyfive{Band 9}, it is likely the best way to tightly constrain the origin of the dust extension in HH212.

All that said, given that the dust extension is co-spatial with the outflow cavity \replytwo{region} as traced by, in particular, \reply{SO, SO$_2$ and }C$^{34}$S, and that the spectral index is significantly different than in the disk \replytwo{\citep{Linetal2021}}, we are inclined to conclude that \reply{a significant portion of} the dust extension is actually \replytwo{associated with the outflow} versus infalling or envelope dust.  \replytwo{Still, without a clearly delineated gradient in $\beta$ between envelope and wind, we cannot rule out that the observed dust extension instead originates from the envelope rather than the wind.}

\replytwo{If} the extension is dust in the wind, then we estimate that the outflow is transporting a non-negligible amount of mass in dust (see Table \ref{tab:mass} and Fig.\ \ref{fig:mass_histogram}, in particular for $>3\sigma$). With our most conservative estimates (i.e.\ using \citealt{OssenkopfHenning1994} opacities and a $>5\sigma$ cut)\reply{, however}, we derive masses of \reply{$\sim 2.76 \pm 0.67 M_{\rm Jup}$ in \replyfive{Band} 9 and $\sim 2.36 \pm 0.20 M_{\rm Jup}$ in \replyfive{Band} 7, corresponding to $750 - 877 M_\oplus$}. Given an adopted disk mass of $0.05M_\odot$, the amount of dust in the wind region likely corresponds to \reply{$\sim4.5 - 5.3\%$} of the total disk mass (Table \ref{tab:mass}).  A larger fraction of mass is also possible if other reasonable assumptions are made (Table \ref{tab:mass}), which in several cases are comparable to the 40\% of the system mass, using Band 3 and 6 observations, measured for L1551 IRS5 by \citet{Sabatini2025}.  While we expect that the lower fraction of mass is most likely for HH212, that may also be representative of its relatively younger evolutionary stage compared with L1551 IRS5.  \replyfour{Our} measurements furthermore represent only one snapshot in time. If the outflow is removing this amount of mass regularly, e.g., \replytwo{on 1000 yr timescales, then over $2\times 10^4$} yr an amount of mass on the order of the current disk mass could be lost from the HH212 system, based on \replytwo{our} estimates from \replyfive{Bands} 7 and 9. \replytwo{Of course, without measurements of the velocity of any dust, this estimate is unfortunately just speculation.}

\reply{Following the methodology of \citet{Tabone20}, who found an ejected gas mass flux from the disk wind (DW) of  
\replytwo{$\dot{M}_{DW}\simeq1-10\times10^{-6}$M$_{\odot}$\ yr$^{-1}$} (scaled to a distance of 400 pc, it would be less by about 10\%), we calculated the dust mass flux.  Granted, there are several sources of uncertainty in the mass flux, which, in the case of the gas, is reported as at least a factor of 10 due to the uncertainty in SO abundance  \citep[according to][]{Podio2015_SO,Lee2018b,Tabone20}.  One also needs to adopt a characteristic velocity of the material, for which there is no assessment for dust.  If we assume that the dust velocity follows the gas velocity, and adopt a velocity ($v_z$) of 5 km s$^{-1}$, as in the models of \citet{Lee2018b}, then we calculate a dust mass flux of \replytwo{$2.5\times10^{-5}\left(\frac{M_{\rm dust}}{1 \textrm{M}_{Jup}}\right)$\ M$_{\odot}$\ yr$^{-1}$}. We note that \citet{Tabone20} adopted $v_z=10$\ km s$^{-1}$, which would scale the mass flux by a factor of 2 larger.  For a dust mass of \replytwo{$\sim$1M$_{Jup}$}, the mass flux is comparable to the upper range of the SO gas flux, and smaller dust masses of \replytwo{$\sim$0.2M$_{Jup}$} would also be consistent.  On the other hand, larger dust masses are likely inconsistent,} \replytwo{as that would imply an unexpectedly low ratio of gas-to-dust mass fluxes (noting that the canonical ISM gas-to-dust mass ratio is generally 100).  We further reiterate that the dust we measure, especially in Region 3, may correspond to a combination of dust launched by the disk wind and entrained envelope material, and moreover the precise gas-to-dust ratios in such specific locations as the disk-wind-envelope interfaces cannot be tightly constrained with the present observations.} 

\reply{Taking into account either the amount of mass carried away in dust over the period of up to \replytwo{$2\times 10^4$ yr, or the estimated} mass flux,} \replytwo{we present a scenario that could} have considerable consequences for the mass budget of any future (or ongoing) planet formation in the disk of HH212. If dust in the wind of HH212 is confirmed, and/or is discovered in progressively more protostellar outflows, then it would demonstrate that outflows not only remove significant amounts of angular momentum, but they can also remove significant amounts of dust, which will have consequences for disk structure and planet formation.

\section{Summary \& Conclusions}
\label{sec:conclude}

\replythree{We present observations to support the case of dust being lifted from the accretion disk by the outflow in HH212.} The spectral index indicates that there are small dust grains in the region just outside the disk but inside the outflow cavity walls, and our mass calculations indicate that this amount of dust is small but significant relative to the mass of the disk.

The entirety of the dust emission beyond the disk is not entirely co-spatial with the outflow cavity, but instead some is along the disk major axis, which suggests that in addition to dust in the outflow, some of the dust may be infalling from the envelope to the outer disk.  The source is clearly in an active stage of early star formation.

One might ask why dust extensions like that presented here are seen in so few sources\replytwo{, either in the literature or as a result of our systematic archival search}. It is evident that these features are faint, and may simply be overlooked in the data, or hidden in the noise.  Indeed, our analysis is based on collecting archival data from five distinct projects \replytwo{that studied the same source, and used} publicly available in the ASA. \replytwo{The present analysis also depended on the availability of compatible observations across multiple ALMA frequency bands, including at least one high-frequency band.}  Additionally, (sub-)millimeter observations preferentially sample dust continuum emission from larger dust grains, but larger, more massive grains are less likely to get carried away by a wind. Far-infrared observations would be more effective for observing emission from smaller dust grains, yet sources that are Class I and younger are generally highly extincted at such wavelengths on scales of the wind launching region; on the other hand, scattered light from Class II sources has been successfully observed at infrared wavelengths \citep[e.g.][]{benistyetal2022_ppvii}, but disk winds are generally weaker at this later stage.

Further observations are needed to confirm the observational detection of dust in the wind of HH212, and to search for similar evidence in other sources.  \reply{While HH212 was selected for our present investigation, it does not necessarily mean that this is a unique source for displacing dust in a disk wind, although it does have an edge-on disk which may be a favorable orientation for noticing the dust extension.  More importantly, this was a source that had conveniently been targeted in five projects (presented here) across multiple ALMA bands, including Band 9, and with sufficient resolution and sensitivity to probe the interface near the disk, envelope and wind; it can serve as a template case for similar studies of other protostars.} 

\begin{acknowledgments}
The National Radio Astronomy Observatory (NRAO) is a facility of the National Science Foundation operated under cooperative agreement by Associated Universities, Inc. This paper makes use of the following ALMA data: ADS/JAO.ALMA\#2012.1.00122.S, ADS/JAO.ALMA\#2015.1.00024.S,\\ ADS/JAO.ALMA\#2016.1.01475.S,\\ ADS/JAO.ALMA\#2017.1.00044.S,\\ ADS/JAO.ALMA\#2017.1.00712.S.\\  ALMA is a partnership of ESO (representing its member states), NSF (USA) and NINS (Japan), together with NRC (Canada), MOST and ASIAA (Taiwan), and KASI (Republic of Korea), in cooperation with the Republic of Chile. The Joint ALMA Observatory is operated by ESO, AUI/NRAO and NAOJ.
This project was completed as part of a student internship program at NRAO. The authors would like to thank the student group at the Chalmers University of Technology: Johanna Matero, Madeleine Karlsson, Viktor Martvall, Aj\v{s}a \'{C}uprija and Veronica Jannisa, as well as Daniel Lin at the University of Virginia, for their work in identifying protostellar sources for this project. Daniel Lin shared his continuum model images of the HH212 disk, as well as extensive knowledge of other aspects of the source. Zhi-Yun Li provided valuable insights at various stages.  The authors would also like to thank the data analysts at the NRAO for their time, expertise, and support, and the Virginia Initiative on Cosmic Origins (VICO) for financial support through the summer student program. The authors acknowledge support from the NASA Astrophysics Theory Program under grant no.\ 80NSSC20K0533. JPR also acknowledges support from the National Science Foundation (NSF) under grant nos.\ AST-1910106 and AST-1910675. PB acknowledges the support of the Swedish Research Council (VR) through contract 2017-04924. 
\end{acknowledgments}

\software{CASA \citep{McMullingetal2007_CASA,CASA2022}, astropy \citep{Astropy2013_I,Astropy2018_II}, matplotlib \citep{Hunter2007_matplotlib}, jupyter \citep{Kluyveretal2016_jupyter}, pandas \citep{pandas}, optool \citep{optool2021ascl}.}

\appendix
\renewcommand{\thetable}{A\arabic{table}}
\setcounter{table}{0}


\section{Archival data sets used for imaging HH212 continuum} \label{appendix:band7}

In this brief section, we \reply{first include a table with archival data sets used in this work (Table \ref{tab:observational_parameters}), and then (Table \ref{tab:band7combination})} present the five data sets from four projects that were incorporated in order to generate a single Band 7 image with sub-arcsecond resolution, as shown in Figure \ref{fig:B7continuum}. Imaging of the Band 7 data was performed in a consistent way to that described in \S \ref{sub:reimaging}.  

\begin{table}[h!]
  \begin{center}
    \caption{Archival data sets used in this work}
    \begin{tabular}{cccccp{5.5cm}}
    \hline
    \hline
      \textbf{Project ID}& \textbf{Band} &  \textbf{Date} & \textbf{MRS} & \textbf{Freq. Range}  & \textbf{Primary References} \\ %
        &   &  & (arcsec) & (GHz) & \\
     \hline
     2017.1.00712.S& 3 & 2017 October 6 & 0.876 & 97.11 - 112.94  & \citet{Linetal2021} \\
     2017.1.00712.S& 6 & 2017 October 4 & 0.440 & 216.82 - 234.94 &  \citet{Linetal2021} \\
     2017.1.00712.S& 6 & 2017 December 31 & 2.778 & 216.82 - 234.94 &  \citet{Linetal2021} \\
     2015.1.00024.S & 7 & 2015 November 5 & 0.357 & 345.68 - 358.93 &  \citet{Linetal2021, Lee2017SciAdv} \\ 
     2016.1.01475.S & 7 & 2016 October 6 & 1.173 & 334.09 - 347.36 &\citet{Taboneetal2017_SO_SO2,Bianchietal2017_HH212methanol}  \\ 
     2016.1.01475.S& 7 & 2016 November 11 & 3.160 & 334.09 - 347.36 &  \citet{Taboneetal2017_SO_SO2,Bianchietal2017_HH212methanol} \\
     2012.1.00122.S & 7 & 2015 August 29 & 1.735 & 345.56 - 358.94 &  \citet{Linetal2021, Lee2017SciAdv} \\
     2017.1.00044.S & 7 & 2017 November 27 & 0.746 & 333.69 - 349.41 & \citet{Linetal2021,Lee_2019} \\
     2012.1.00122.S& 9 &  2015 July 26 & 0.772 & 688.10 - 694.76 & \citet{Linetal2021, Lee2017SciAdv}\\
      \hline
    \end{tabular}
    \label{tab:observational_parameters}
    \tablecomments{MRS is the ``Maximum recoverable scale'' reported in the ALMA Science Archive (ASA). Frequency range also corresponds to values reported in the ASA. References given are those that initially analyze and present the continuum data from the corresponding programs.}
  \end{center}
\end{table}

\begin{table}[ht]
\setlength{\tabcolsep}{3pt}
  \begin{center}
    \caption{Band 7 Data and Imaging Parameters}
    \label{tab:band7combination}
    \begin{tabular}{ccccccc}
    \hline
    \hline
      \textbf{Project ID} &  \textbf{\replyfive{$\theta_{beam}$\ (Maj.$\times$Min.) }} & \textbf{PA} & \textbf{Cell Size} & \textbf{Peak Intensity} & \textbf{Total Flux} & \textbf{$\sigma$} \\ %
      \hline
        &  [\arcsec]  & [\degree] & [\arcsec] &[Jy beam$^{-1}$]&[Jy] & [mJy beam$^{-1}$]\\
     \hline
      2012.1.00122.S & 0.1996 x 0.1346 & 85.46 & 0.035 & 0.1 & 0.8 & 0.2  \\
      2015.1.00024.S & 0.0205 x 0.0185 & -71.25 & 0.0036 & 0.003 & 0.08 & 0.2  \\
      2016.1.01475.S(x259) & 0.0963 x 0.0922 & -46.67 & 0.019 & 0.03 & 0.3 & 0.2  \\
      2016.1.01475.S(x25b) & 0.3912 x 0.2825 & -69.54 & 0.0075 & 0.1 & 0.4 & 0.4  \\
      2017.1.00044.S & 0.0365 x 0.0331 & -81.62 & 0.0065 & 0.009 & 0.2 & 0.04  \\
      \hline
    \end{tabular}
    \tablecomments{Observational parameters correspond to projects, and their respective Member Obs Unit Sets (MOUS) as indicated by the final 4 characters in parentheses where applicable, that observed Band 7 and were combined in this work to form the 345 GHz continuum image in Figure \ref{fig:B7continuum}.}
  \end{center}
\end{table}

\replyfour{\section{Regions defined for calculation }\label{appendix:regionsdefined}
In order to assess spectral index and quantify dust mass, we partitioned the maps into three regions which we describe here and tabulate in Table \ref{tab:region_definitions}.  Region 1 corresponds to the entire field of view, including the compact, central, disk-dominated component, and it is excluded from the physical interpretation of $\alpha$\ and $\beta$ because continuum in the HH212 disk is known to be optically thick in Band 7 and higher frequencies \citep{Linetal2021,Hu2025}. Region 2 contains extended continuum emission outside of a mask encompassing significant disk contributions; it may include emission arising from an extended envelope along the disk major axis.  Region 3 is defined as a subset of Region 2 that spatially coincides with the cavity/wind structure traced by SO and bounded by C$^{34}$S emission.  It therefore represents the part of the extended dust emission most plausibly associated with the molecular outflow cavity.  Thus, Region 3 is nested within Region 2, while Region 1 is treated separately because of optical-depth effects.  These definitions are operational rather than purely physical, and they allow us to distinguish the regions of dust continuum emission that are disk-dominated, extended non-disk, and extended emission associated with the outflow cavity. Table \ref{tab:region_definitions} defines the pixels that are included and excluded from each region.}

\begin{table*}
\centering
\caption{Operational definitions of analysis regions}
\label{tab:region_definitions}
\begin{tabular}{c p{0.36\textwidth} p{0.36\textwidth}}
\hline
Region & Included pixels & Excluded pixels \\
\hline
1 &
Pixels for which both the Band 7 and Band 9 continuum intensities are above $3-5\sigma$. &
Pixels for which either Band 7 or Band 9 continuum intensity is below $3-5\sigma$. \\
2 &
All pixels in Region~1, plus pixels outside the central ellipse, where the central ellipse is defined to encompass the $4\sigma$ level of the two-dimensional Gaussian fit to the Band 9 continuum emission. &
All pixels excluded from Region~1, plus pixels inside the central ellipse defined by the $4\sigma$ level of the two-dimensional Gaussian fit to the Band 9 continuum emission. \\
3 &
All pixels in Region~2, plus pixels inside the green ellipse that encloses the southern/southwestern area where significant C$^{34}$S and SO emission is observed. &
All pixels excluded from Region~2, plus pixels outside the green ellipse. \\
\hline
\end{tabular}
\end{table*}

\section{Archival data sets to demonstrate molecular line emission} \label{appendix:othermolecules}
\reply{Archival data exist for several molecular lines observed towards HH212 that pertain to the disk wind and cavity. The maps we show in the following figures have the same spatial scales and annotations as in Figures \ref{fig:C34S} and \ref{fig:contmasking}, in order to make a qualitative comparison between outflow and envelope structures.  Data shown in this Appendix are those presented by \citet{Taboneetal2017_SO_SO2}}

\begin{figure}[ht]
\centering
\includegraphics[width=0.48\textwidth]{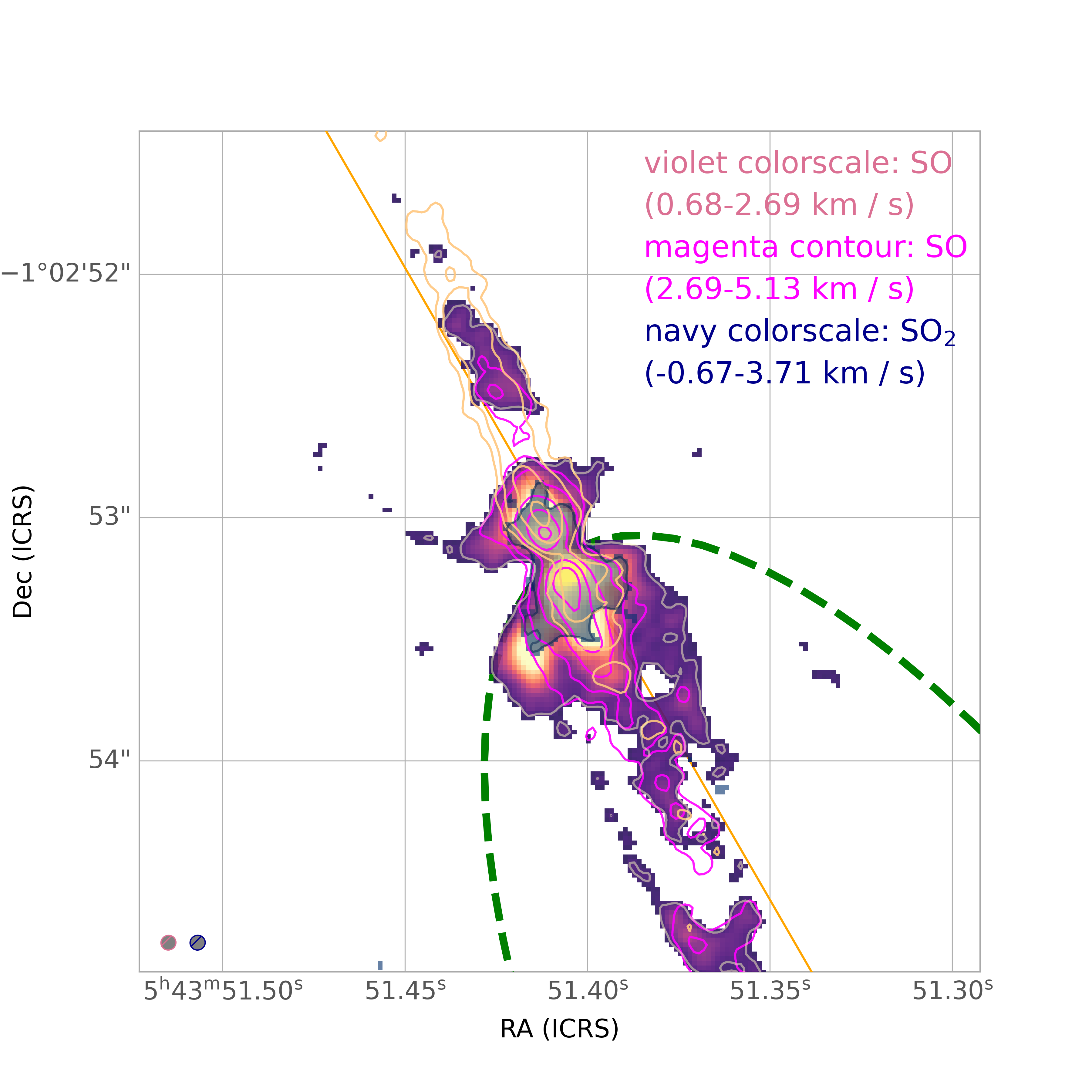}
\includegraphics[width=0.48\textwidth]{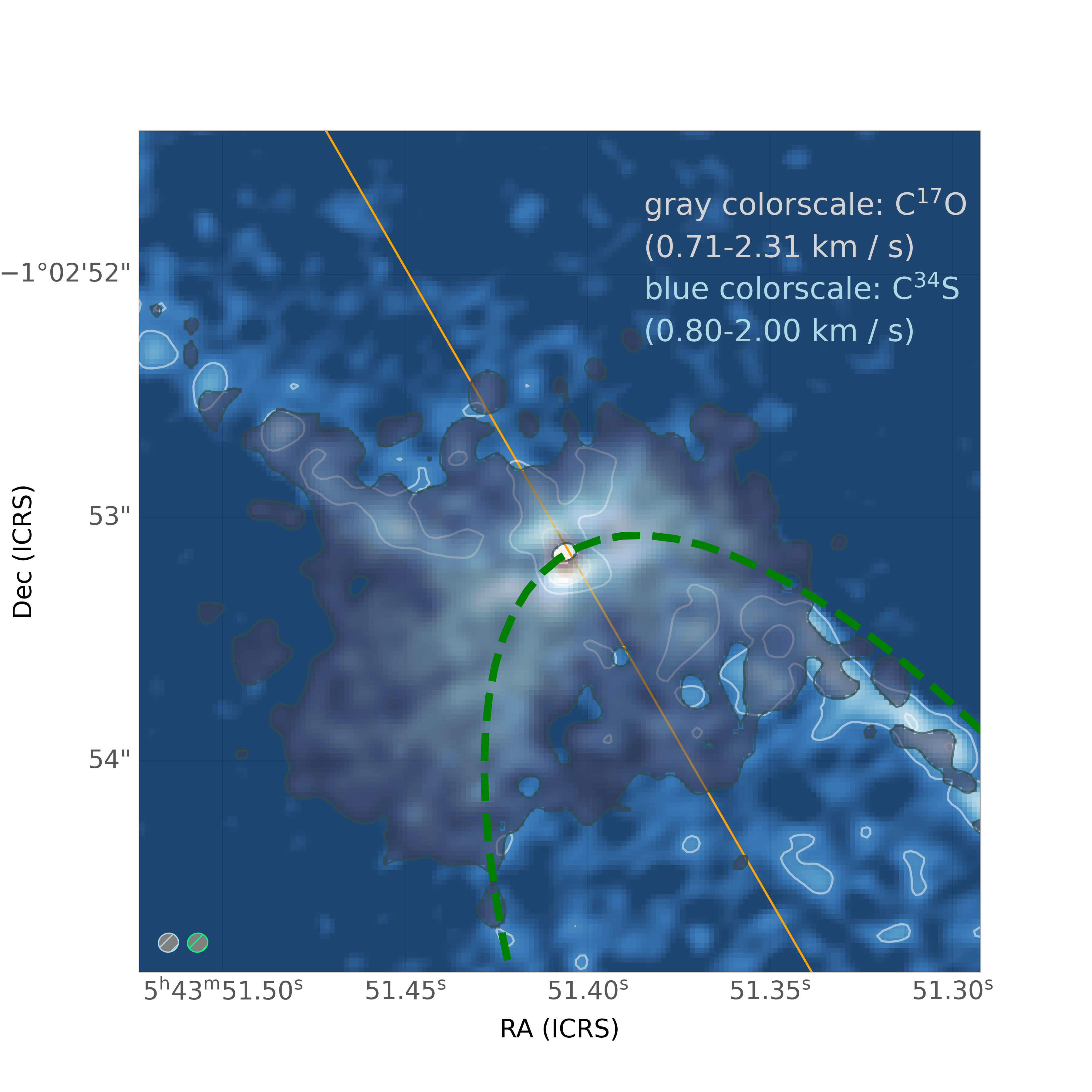}
\caption{
(left) SO (9$_8$-8$_7$) and SO$_2$ (8$_{2,6}$-7$_{1,7}$) integrated emission moment 0 maps using data from \cite{Taboneetal2017_SO_SO2} (magenta and blue color scales, respectively); both SO and SO$_2$ are masked below 2.5 times the rms noise level, and the outer contours signify the $3\sigma$ level of the respective moment maps. \replytwo{Magenta contours (at levels of 3, 5, 10, 15$\sigma$) show the higher-velocity redshifted channels of SO (note that the full range of higher-velocity redshifted channels were not observed).} (right) C$^{17}$O (3-2) and C$^{34}$S \citep{Taboneetal2017_SO_SO2} integrated emission moment 0 maps (\replytwo{gray} and blue color scales, respectively); C$^{17}$O is masked below 5 times the rms noise level of the moment 0 map, and the outer contour signifies the $5\sigma$ level. In both panels, the dashed green ellipse and orange diagonal lines are the same as in Figure \ref{fig:C34S}, for reference. Small ellipses in lower left corners of each panel represent the synthesized beams of the respective datasets indicated with the same colors in the upper right corners.}
\label{fig:appendix_soso2}
\end{figure}

\reply{Figure \ref{fig:appendix_soso2} shows low-velocity emission from SO and SO$_2$ \citep[see also][their Fig.\ 4]{Lee2018b}. These coincide spatially, with SO being slightly more extended towards the south.  \replytwo{We also show emission from higher-velocity red-shifted SO channels, where we see emission that is slightly more collimated. We also note that these velocities are the line-of-sight velocities without any correction for inclination angle of the outflow.} Both SO and SO$_2$ emit from within the cavity structure outlined by C$^{34}$S \replyfour{\citep[see also][who show SO in the context of SiO and CO]{Lopezvazquez2024}}.  
We argue that \replyfour{SO and SO$_2$ line emission morphologies near the outflow base} are qualitatively similar to the dust extension we observe in Band 9 at the south edge of the disk (``Region 3''; Fig.\ \ref{fig:C34S}).}

We also inspected the envelope structure, as traced by C$^{17}$O (J=3-2). Figure \ref{fig:appendix_soso2} (right) shows the C$^{17}$O envelope structure, with C$^{34}$S in the background for context \citep[see also][]{Codellaetal2014_HH212}.  One can see that the envelope comprises extended emission in all directions around the disk, on scales of $\gtrsim0.5\arcsec$.  The C$^{17}$O envelope reflects some correspondence with the cavity as traced by the low-velocity component of C$^{34}$S, as well as infall and rotation on smaller scales near the disk.

\section{Methodology for directional determination} \label{appendix:direction}
\replythree{Residual structures in the Bands 7 and 9 continuum maps are clearly detected above the $>$3$\sigma$ level, but to determine whether they are oriented in a preferred direction rather than arising from random fluctuations, we performed a dedicated quantitative analysis. The procedure is as follows.

First, each continuum map was fitted with a two-dimensional Gaussian profile using the \texttt{astropy.modeling} package. The fit provides the source centroid, orientation angle, and full width at half maximum (FWHM) along the major and minor axes. An exclusion ellipse was then defined as being twice as large as the fitted FWHM, and residual emission was sampled within an annular region extending a factor of five beyond this ellipse. This ensures that only emission outside the central compact source contributes to the directional test.

Within this annular region, the residual intensity was divided into 21 angular bins (corresponding to a bin width of $\sim$17°). Angles are defined relative to north, increasing eastward, following the standard astronomical convention. For each bin, we computed the mean residual intensity.

To evaluate whether the emission exhibits a statistically preferred orientation, we applied a rolling-sum analysis combined with a permutation test. In this approach, contiguous sets of bins are summed, and the strongest observed cluster (of any width) is compared against a large ensemble of randomized realizations in which the bin intensities are permuted (10,000 permutations). The resulting p-value quantifies the probability that a residual excess as strong as the observed one could arise purely by chance if the emission were isotropically distributed in angle. A low p-value therefore indicates a statistically significant directional preference.

For Band 7, the most significant residual sector spans 6 contiguous bins, oriented from –171° to –81° with respect to north, with a p-value of 0.0014. A second, non-overlapping sector was not significant (p = $\simeq$1). For Band 9, the most significant residual sector spans 6 bins, oriented from –153° to –63°, with a p-value of 0.0006. The second best sector was not significant (p $\simeq$1).

In both bands, therefore, the analysis identifies a statistically significant residual excess oriented toward the south–east, in the same direction as the southern outflow launching region indicated by SO and C$^{34}$S (see Fig. \ref{fig:appendix_soso2}) and discussed in \S \ref{sub:dustmorph}. Given the low p-values, the likelihood that this alignment results from random fluctuations is small, and this suggest a genuine preferred direction in the residual emission.}

\begin{figure}[ht]
\centering
\includegraphics[width=1.0\textwidth]{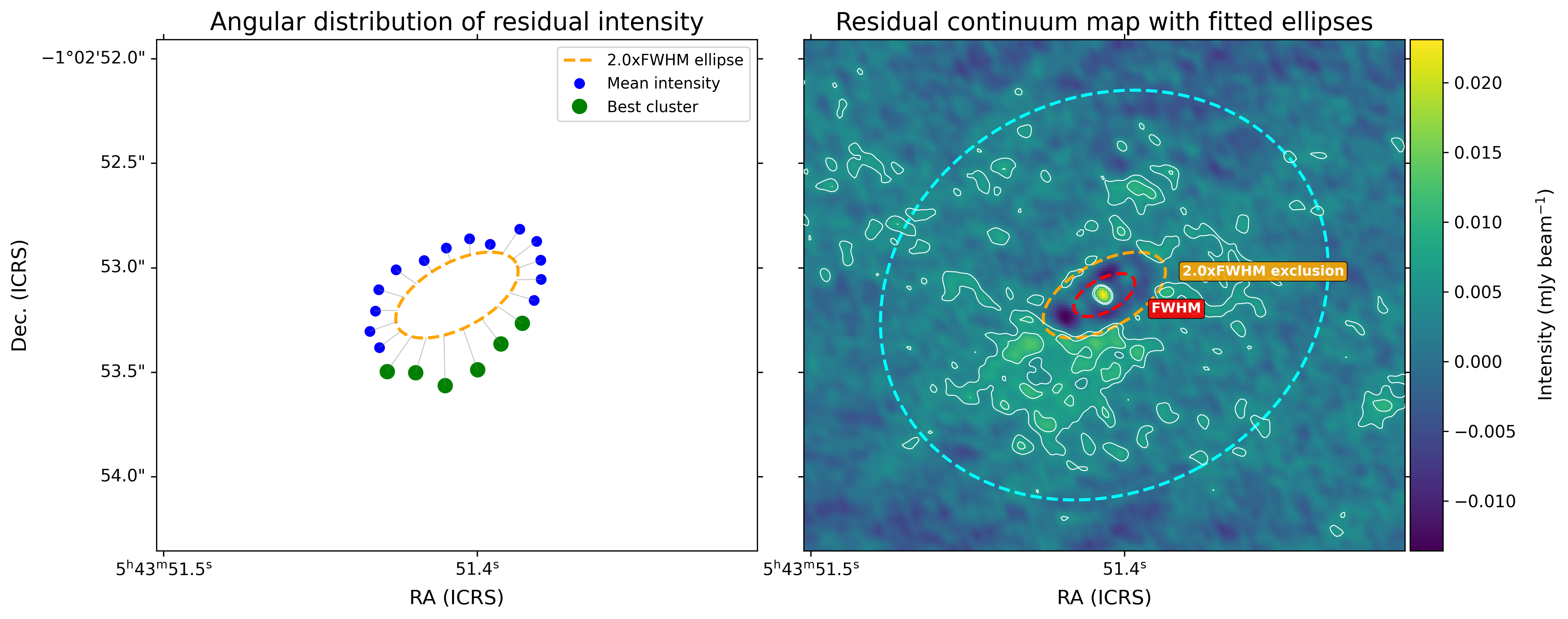}
\caption{\replythree{\textit{Left:} Angular distribution of residual continuum emission in Band 9, averaged in angular bins around the fitted exclusion ellipse. The orange dashed ellipse marks the exclusion region, blue points show the mean intensity, and green points highlight the most significant cluster. \textit{Right:} Residual continuum map after subtraction of the best-fit 2D Gaussian. The red dashed ellipse shows the FWHM, the orange ellipse the exclusion region (2 x FWHM) , and the cyan ellipse the offset boundary used in the diretional analysis.}}
\label{fig:methodology_direction}
\end{figure}
\begin{figure}[ht]
\centering
\includegraphics[width=1.0\textwidth]{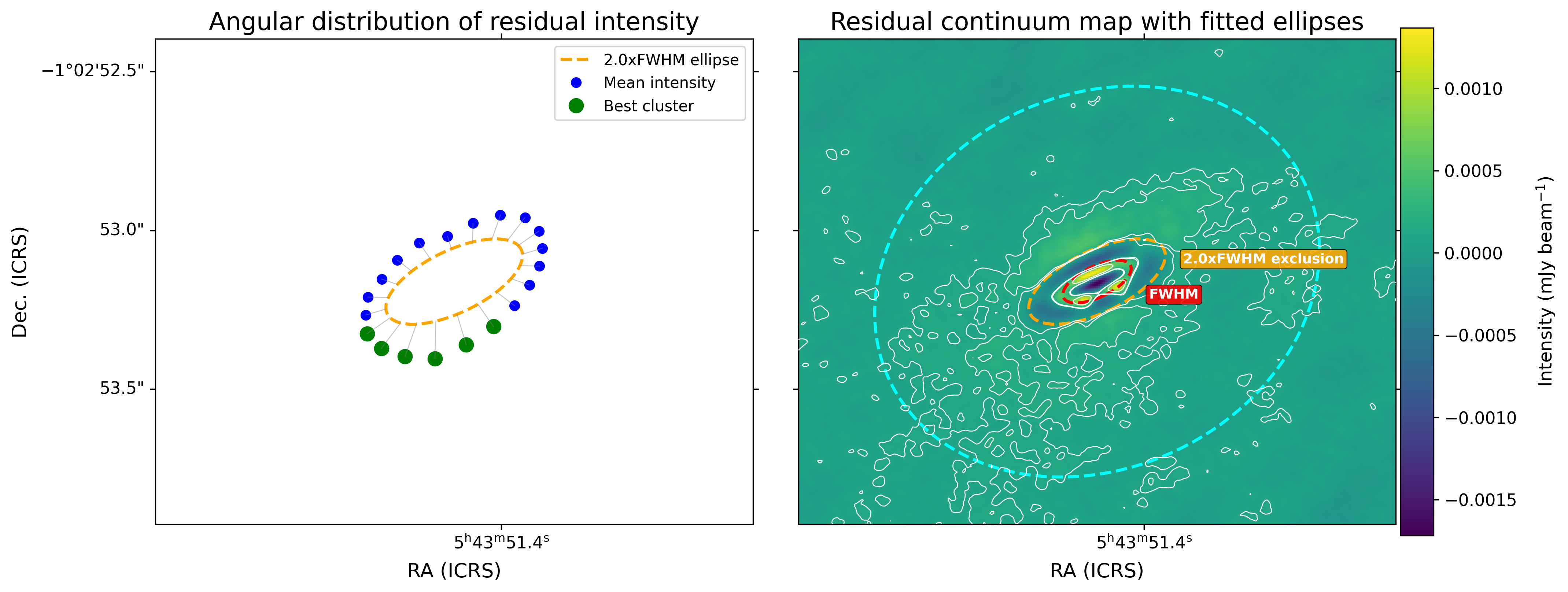}
\caption{\replythree{Same as Fig.~\ref{fig:methodology_direction} but for Band 7.}}
\label{fig:methodology_direction_band7}
\end{figure}



\bibliography{hh212}{}

@ARTICLE{Lopezvazquez2024,
       author = {{L{\'o}pez-V{\'a}zquez}, J.~A. and {Lee}, Chin-Fei and {Shang}, Hsien and {Cabrit}, Sylvie and {Krasnopolsky}, Ruben and {Codella}, Claudio and {Liu}, Chun-Fan and {Podio}, Linda and {Dutta}, Somnath and {Murphy}, A. and {Wiseman}, Jennifer},
        title = "{Multiple Components of the Outflow in the Protostellar System HH 212: Outer Outflow Shell, Rotating Wind, Shocked Wind, and Jet}",
      journal = {\apj},
         year = 2024,
        month = dec,
       volume = {977},
       number = {1},
          eid = {126},
        pages = {126},
          doi = {10.3847/1538-4357/ad8eb4},
archivePrefix = {arXiv},
       eprint = {2411.01728},
 primaryClass = {astro-ph.SR},
       adsurl = {https://ui.adsabs.harvard.edu/abs/2024ApJ...977..126L}
}

@ARTICLE{Hu2025,
       author = {{Hu}, Ying-Chi and {Lee}, Chin-Fei and {Lin}, Zhe-Yu Daniel and {Li}, Zhi-Yun and {Tobin}, John J. and {Lai}, Shih-Ping},
        title = "{Early Grain Growth in the Young Protostellar Disk HH 212 Supported by Dust Self-scattering Modeling}",
      journal = {\apj},
         year = 2025,
        month = mar,
       volume = {981},
       number = {1},
          eid = {12},
        pages = {12},
          doi = {10.3847/1538-4357/adabe1},
archivePrefix = {arXiv},
       eprint = {2412.00305},
 primaryClass = {astro-ph.SR},
       adsurl = {https://ui.adsabs.harvard.edu/abs/2025ApJ...981...12H}
}

@ARTICLE{Sabatini2025,
       author = {{Sabatini}, G. and {Bianchi}, E. and {Chandler}, C.~J. and {Cacciapuoti}, L. and {Podio}, L. and {Maureira}, M.~J. and {Codella}, C. and {Ceccarelli}, C. and {Sakai}, N. and {Testi}, L. and {Toci}, C. and {Svoboda}, B. and {Sakai}, T. and {Bouvier}, M. and {Caselli}, P. and {Cuello}, N. and {De Simone}, M. and {J{\'\i}menez-Serra}, I. and {Johnstone}, D. and {Loinard}, L. and {Zhang}, Z.~E. and {Yamamoto}, S.},
        title = "{FAUST: XXIV. Large dust grains in the protostellar outflow cavity walls of the Class I binary L1551 IRS5}",
      journal = {\aap},
         year = 2025,
        month = jun,
       volume = {698},
          eid = {L16},
        pages = {L16},
          doi = {10.1051/0004-6361/202554750},
archivePrefix = {arXiv},
       eprint = {2505.13596},
 primaryClass = {astro-ph.GA},
       adsurl = {https://ui.adsabs.harvard.edu/abs/2025A&A...698L..16S}
}

@ARTICLE{BoothClarke2021_dusttransport,
       author = {{Booth}, Richard A. and {Clarke}, Cathie J.},
        title = "{Modelling the delivery of dust from discs to ionized winds}",
      journal = {\mnras},
         year = 2021,
        month = apr,
       volume = {502},
       number = {2},
        pages = {1569-1578},
          doi = {10.1093/mnras/stab090},
archivePrefix = {arXiv},
       eprint = {2101.04121},
 primaryClass = {astro-ph.EP},
       adsurl = {https://ui.adsabs.harvard.edu/abs/2021MNRAS.502.1569B}
}

@ARTICLE{Giacaloneetal2019_dusttransport,
       author = {{Giacalone}, Steven and {Teitler}, Seth and {K{\"o}nigl}, Arieh and {Krijt}, Sebastiaan and {Ciesla}, Fred J.},
        title = "{Dust Transport and Processing in Centrifugally Driven Protoplanetary Disk Winds}",
      journal = {\apj},
         year = 2019,
        month = sep,
       volume = {882},
       number = {1},
          eid = {33},
        pages = {33},
          doi = {10.3847/1538-4357/ab311a},
archivePrefix = {arXiv},
       eprint = {1907.04961},
 primaryClass = {astro-ph.SR},
       adsurl = {https://ui.adsabs.harvard.edu/abs/2019ApJ...882...33G}
}

@ARTICLE{Franzetal2022a_imaging,
       author = {{Franz}, R. and {Ercolano}, B. and {Casassus}, S. and {Picogna}, G. and {Birnstiel}, T. and {P{\'e}rez}, S. and {Rab}, Ch. and {Sharma}, A.},
        title = "{Dust entrainment in photoevaporative winds: Densities and imaging}",
      journal = {\aap},
         year = 2022,
        month = jan,
       volume = {657},
          eid = {A69},
        pages = {A69},
          doi = {10.1051/0004-6361/202140812},
archivePrefix = {arXiv},
       eprint = {2110.10637},
 primaryClass = {astro-ph.EP},
       adsurl = {https://ui.adsabs.harvard.edu/abs/2022A&A...657A..69F}
}

@ARTICLE{Franzetal2020_dustphotoevapI,
       author = {{Franz}, R. and {Picogna}, G. and {Ercolano}, B. and {Birnstiel}, T.},
        title = "{Dust entrainment in photoevaporative winds: The impact of X-rays}",
      journal = {\aap},
         year = 2020,
        month = mar,
       volume = {635},
          eid = {A53},
        pages = {A53},
          doi = {10.1051/0004-6361/201936615},
archivePrefix = {arXiv},
       eprint = {2001.10545},
 primaryClass = {astro-ph.EP},
       adsurl = {https://ui.adsabs.harvard.edu/abs/2020A&A...635A..53F}
}

@ARTICLE{Beckwithetal1990,
       author = {{Beckwith}, Steven V.~W. and {Sargent}, Anneila I. and {Chini}, Rolf S. and {Guesten}, Rolf},
        title = "{A Survey for Circumstellar Disks around Young Stellar Objects}",
      journal = {\aj},
         year = 1990,
        month = mar,
       volume = {99},
        pages = {924},
          doi = {10.1086/115385},
       adsurl = {https://ui.adsabs.harvard.edu/abs/1990AJ.....99..924B}
}

@ARTICLE{Riccietal2010,
       author = {{Ricci}, L. and {Testi}, L. and {Natta}, A. and {Neri}, R. and {Cabrit}, S. and {Herczeg}, G.~J.},
        title = "{Dust properties of protoplanetary disks in the Taurus-Auriga star forming region from millimeter wavelengths}",
      journal = {\aap},
         year = 2010,
        month = mar,
       volume = {512},
          eid = {A15},
        pages = {A15},
          doi = {10.1051/0004-6361/200913403},
archivePrefix = {arXiv},
       eprint = {0912.3356},
 primaryClass = {astro-ph.EP},
       adsurl = {https://ui.adsabs.harvard.edu/abs/2010A&A...512A..15R}
}

@ARTICLE{Juvela2015,
       author = {{Juvela}, M. and {Demyk}, K. and {Doi}, Y. and {Hughes}, A. and {Lef{\`e}vre}, C. and {Marshall}, D.~J. and {Meny}, C. and {Montillaud}, J. and {Pagani}, L. and {Paradis}, D. and {Ristorcelli}, I. and {Malinen}, J. and {Montier}, L.~A. and {Paladini}, R. and {Pelkonen}, V. -M. and {Rivera-Ingraham}, A.},
        title = "{Galactic cold cores. VI. Dust opacity spectral index}",
      journal = {\aap},
         year = 2015,
        month = dec,
       volume = {584},
          eid = {A94},
        pages = {A94},
          doi = {10.1051/0004-6361/201425269},
archivePrefix = {arXiv},
       eprint = {1509.08023},
 primaryClass = {astro-ph.GA},
       adsurl = {https://ui.adsabs.harvard.edu/abs/2015A&A...584A..94J}
}

@ARTICLE{miotello2014,
       author = {{Miotello}, A. and {Testi}, L. and {Lodato}, G. and {Ricci}, L. and {Rosotti}, G. and {Brooks}, K. and {Maury}, A. and {Natta}, A.},
        title = "{Grain growth in the envelopes and disks of Class I protostars}",
      journal = {\aap},
         year = 2014,
        month = jul,
       volume = {567},
          eid = {A32},
        pages = {A32},
          doi = {10.1051/0004-6361/201322945},
archivePrefix = {arXiv},
       eprint = {1405.0821},
 primaryClass = {astro-ph.SR},
       adsurl = {https://ui.adsabs.harvard.edu/abs/2014A&A...567A..32M}
}

@ARTICLE{bracco2017,
       author = {{Bracco}, A. and {Palmeirim}, P. and {Andr{\'e}}, Ph. and {Adam}, R. and {Ade}, P. and {Bacmann}, A. and {Beelen}, A. and {Beno{\^\i}t}, A. and {Bideaud}, A. and {Billot}, N. and {Bourrion}, O. and {Calvo}, M. and {Catalano}, A. and {Coiffard}, G. and {Comis}, B. and {D'Addabbo}, A. and {D{\'e}sert}, F. -X. and {Didelon}, P. and {Doyle}, S. and {Goupy}, J. and {K{\"o}nyves}, V. and {Kramer}, C. and {Lagache}, G. and {Leclercq}, S. and {Mac{\'\i}as-P{\'e}rez}, J.~F. and {Maury}, A. and {Mauskopf}, P. and {Mayet}, F. and {Monfardini}, A. and {Motte}, F. and {Pajot}, F. and {Pascale}, E. and {Peretto}, N. and {Perotto}, L. and {Pisano}, G. and {Ponthieu}, N. and {Rev{\'e}ret}, V. and {Rigby}, A. and {Ritacco}, A. and {Rodriguez}, L. and {Romero}, C. and {Roy}, A. and {Ruppin}, F. and {Schuster}, K. and {Sievers}, A. and {Triqueneaux}, S. and {Tucker}, C. and {Zylka}, R.},
        title = "{Probing changes of dust properties along a chain of solar-type prestellar and protostellar cores in Taurus with NIKA}",
      journal = {\aap},
         year = 2017,
        month = aug,
       volume = {604},
          eid = {A52},
        pages = {A52},
          doi = {10.1051/0004-6361/201731117},
archivePrefix = {arXiv},
       eprint = {1706.08407},
 primaryClass = {astro-ph.GA},
       adsurl = {https://ui.adsabs.harvard.edu/abs/2017A&A...604A..52B}
}

@ARTICLE{Linetal2021,
       author = {{Lin}, Zhe-Yu Daniel and {Lee}, Chin-Fei and {Li}, Zhi-Yun and {Tobin}, John J. and {Turner}, Neal J.},
        title = "{Inferring (sub)millimetre dust opacities and temperature structure in edge-on protostellar discs from resolved multiwavelength continuum observations: the case of the HH 212 disc}",
      journal = {\mnras},
         year = 2021,
        month = feb,
       volume = {501},
       number = {1},
        pages = {1316-1335},
          doi = {10.1093/mnras/staa3685},
archivePrefix = {arXiv},
       eprint = {2008.08627},
 primaryClass = {astro-ph.SR},
       adsurl = {https://ui.adsabs.harvard.edu/abs/2021MNRAS.501.1316L}
}

@ARTICLE{Lee2014,
       author = {{Lee}, Chin-Fei and {Hirano}, Naomi and {Zhang}, Qizhou and {Shang}, Hsien and {Ho}, Paul T.~P. and {Krasnopolsky}, Ruben},
        title = "{ALMA Results of the Pseudodisk, Rotating Disk, and Jet in the Continuum and HCO$^{+}$ in the Protostellar System HH 212}",
      journal = {\apj},
         year = 2014,
        month = may,
       volume = {786},
       number = {2},
          eid = {114},
        pages = {114},
          doi = {10.1088/0004-637X/786/2/114},
archivePrefix = {arXiv},
       eprint = {1403.5853},
 primaryClass = {astro-ph.GA},
       adsurl = {https://ui.adsabs.harvard.edu/abs/2014ApJ...786..114L}
}

@ARTICLE{Lee2017NatAs,
       author = {{Lee}, Chin-Fei and {Ho}, Paul. T.~P. and {Li}, Zhi-Yun and {Hirano}, Naomi and {Zhang}, Qizhou and {Shang}, Hsien},
        title = "{A rotating protostellar jet launched from the innermost disk of HH 212}",
      journal = {Nature Astronomy},
         year = 2017,
        month = jul,
       volume = {1},
          eid = {0152},
        pages = {0152},
          doi = {10.1038/s41550-017-0152},
archivePrefix = {arXiv},
       eprint = {1706.06343},
 primaryClass = {astro-ph.GA},
       adsurl = {https://ui.adsabs.harvard.edu/abs/2017NatAs...1E.152L}
}

@ARTICLE{Lee2017ApJ,
       author = {{Lee}, Chin-Fei and {Li}, Zhi-Yun and {Ho}, Paul T.~P. and {Hirano}, Naomi and {Zhang}, Qizhou and {Shang}, Hsien},
        title = "{Formation and Atmosphere of Complex Organic Molecules of the HH 212 Protostellar Disk}",
      journal = {\apj},
         year = 2017,
        month = jul,
       volume = {843},
       number = {1},
          eid = {27},
        pages = {27},
          doi = {10.3847/1538-4357/aa7757},
archivePrefix = {arXiv},
       eprint = {1706.06041},
 primaryClass = {astro-ph.GA},
       adsurl = {https://ui.adsabs.harvard.edu/abs/2017ApJ...843...27L}
}

@ARTICLE{Lee2017SciAdv,
       author = {{Lee}, Chin-Fei and {Li}, Zhi-Yun and {Ho}, Paul T.~P. and {Hirano}, Naomi and {Zhang}, Qizhou and {Shang}, Hsien},
        title = "{First detection of equatorial dark dust lane in a protostellar disk at submillimeter wavelength}",
      journal = {Science Advances},
         year = 2017,
        month = apr,
       volume = {3},
       number = {4},
          eid = {e1602935},
        pages = {e1602935},
          doi = {10.1126/sciadv.1602935},
archivePrefix = {arXiv},
       eprint = {1704.08962},
 primaryClass = {astro-ph.GA},
       adsurl = {https://ui.adsabs.harvard.edu/abs/2017SciA....3E2935L}
}

@article{Lee_2018a,
   title={ALMA Dust Polarization Observations of Two Young Edge-on Protostellar Disks},
   volume={854},
   ISSN={1538-4357},
   url={http://dx.doi.org/10.3847/1538-4357/aaa769},
   DOI={10.3847/1538-4357/aaa769},
   number={1},
   journal={The Astrophysical Journal},
   publisher={American Astronomical Society},
   author={Lee, Chin-Fei and Li, Zhi-Yun and Ching, Tao-Chung and Lai, Shih-Ping and Yang, Haifeng},
   year={2018},
   month={Feb},
   pages={56}
}

@article{Lee2018b,
	Adsurl = {http://adsabs.harvard.edu/abs/2018ApJ...856...14L},
	Archiveprefix = {arXiv},
	Arxivurl = {http://arxiv.org/abs/1802.03668},
	Author = {{Lee}, C.-F. and {Li}, Z.-Y. and {Codella}, C. and {Ho}, P.~T.~P. and {Podio}, L. and {Hirano}, N. and {Shang}, H. and {Turner}, N.~J. and {Zhang}, Q.},
	Doi = {10.3847/1538-4357/aaae6d},
	Eid = {14},
	Eprint = {1802.03668},
	Journal = {\apj},
	Month = mar,
	Pages = {14},
	Title = {{A 100 au Wide Bipolar Rotating Shell Emanating from the HH 212 Protostellar Disk: A Disk Wind?}},
	Volume = 856,
	Year = 2018,
}

@article{Lee_2019,
   title={First Abundance Measurement of Organic Molecules in the Atmosphere of HH 212 Protostellar Disk},
   volume={876},
   ISSN={1538-4357},
   url={http://dx.doi.org/10.3847/1538-4357/ab15db},
   DOI={10.3847/1538-4357/ab15db},
   number={1},
   journal={The Astrophysical Journal},
   publisher={American Astronomical Society},
   author={Lee, Chin-Fei and Codella, Claudio and Li, Zhi-Yun and Liu, Sheng-Yuan},
   year={2019},
   month={May},
   pages={63}
}

@ARTICLE{LeeTaboneetal2021,
       author = {{Lee}, Chin-Fei and {Tabone}, Benoit and {Cabrit}, Sylvie and {Codella}, Claudio and {Podio}, Linda and {Ferreira}, Jonathan and {Jacquemin-Ide}, Jonatan},
        title = "{First Detection of Interaction between a Magnetic Disk Wind and an Episodic Jet in a Protostellar System}",
      journal = {\apjl},
         year = 2021,
        month = feb,
       volume = {907},
       number = {2},
          eid = {L41},
        pages = {L41},
          doi = {10.3847/2041-8213/abda38},
archivePrefix = {arXiv},
       eprint = {2101.03293},
 primaryClass = {astro-ph.GA},
       adsurl = {https://ui.adsabs.harvard.edu/abs/2021ApJ...907L..41L}
}

@ARTICLE{LeeLietal2021b,
       author = {{Lee}, Chin-Fei and {Li}, Zhi-Yun and {Yang}, Haifeng and {Daniel Lin}, Zhe-Yu and {Ching}, Tao-Chung and {Lai}, Shih-Ping},
        title = "{What Produces Dust Polarization in the HH 212 Protostellar Disk at 878 {\ensuremath{\mu}}m: Dust Self-scattering or Dichroic Extinction?}",
      journal = {\apj},
         year = 2021,
        month = mar,
       volume = {910},
       number = {1},
          eid = {75},
        pages = {75},
          doi = {10.3847/1538-4357/abe53a},
archivePrefix = {arXiv},
       eprint = {2102.11481},
 primaryClass = {astro-ph.GA},
       adsurl = {https://ui.adsabs.harvard.edu/abs/2021ApJ...910...75L}
}

@ARTICLE{LeeLietal2022,
       author = {{Lee}, Chin-Fei and {Li}, Zhi-Yun and {Shang}, Hsien and {Hirano}, Naomi},
        title = "{Magnetocentrifugal Origin for Protostellar Jets Validated through Detection of Radial Flow at the Jet Base}",
      journal = {\apjl},
         year = 2022,
        month = mar,
       volume = {927},
       number = {2},
          eid = {L27},
        pages = {L27},
          doi = {10.3847/2041-8213/ac59c0},
archivePrefix = {arXiv},
       eprint = {2203.00961},
 primaryClass = {astro-ph.GA},
       adsurl = {https://ui.adsabs.harvard.edu/abs/2022ApJ...927L..27L}
}

@ARTICLE{Leeetal2008,
       author = {{Lee}, Chin-Fei and {Ho}, Paul T.~P. and {Bourke}, Tyler L. and {Hirano}, Naomi and {Shang}, Hsien and {Zhang}, Qizhou},
        title = "{SiO Shocks of the Protostellar Jet HH 212: A Search for Jet Rotation}",
      journal = {\apj},
         year = 2008,
        month = oct,
       volume = {685},
       number = {2},
        pages = {1026-1032},
          doi = {10.1086/591177},
       adsurl = {https://ui.adsabs.harvard.edu/abs/2008ApJ...685.1026L}
}

@ARTICLE{Leeetal2006,
       author = {{Lee}, Chin-Fei and {Ho}, Paul T.~P. and {Beuther}, Henrik and {Bourke}, Tyler L. and {Zhang}, Qizhou and {Hirano}, Naomi and {Shang}, Hsien},
        title = "{Infall and Outflow around the HH 212 Protostellar System}",
      journal = {\apj},
         year = 2006,
        month = mar,
       volume = {639},
       number = {1},
        pages = {292-302},
          doi = {10.1086/499297},
archivePrefix = {arXiv},
       eprint = {astro-ph/0511059},
 primaryClass = {astro-ph},
       adsurl = {https://ui.adsabs.harvard.edu/abs/2006ApJ...639..292L}
}

@ARTICLE{Leeetal2015,
       author = {{Lee}, Chin-Fei and {Hirano}, Naomi and {Zhang}, Qizhou and {Shang}, Hsien and {Ho}, Paul T.~P. and {Mizuno}, Yosuke},
        title = "{Jet Motion, Internal Working Surfaces, and Nested Shells in the Protostellar System HH 212}",
      journal = {\apj},
         year = 2015,
        month = jun,
       volume = {805},
       number = {2},
          eid = {186},
        pages = {186},
          doi = {10.1088/0004-637X/805/2/186},
archivePrefix = {arXiv},
       eprint = {1503.07362},
 primaryClass = {astro-ph.GA},
       adsurl = {https://ui.adsabs.harvard.edu/abs/2015ApJ...805..186L}
}

@ARTICLE{Tabone20,
       author = {{Tabone}, B. and {Cabrit}, S. and {Pineau des For{\^e}ts}, G. and {Ferreira}, J. and {Gusdorf}, A. and {Podio}, L. and {Bianchi}, E. and {Chapillon}, E. and {Codella}, C. and {Gueth}, F.},
        title = "{Constraining MHD disk winds with ALMA. Apparent rotation signatures and application to HH212}",
      journal = {\aap},
         year = 2020,
        month = aug,
       volume = {640},
          eid = {A82},
        pages = {A82},
          doi = {10.1051/0004-6361/201834377},
archivePrefix = {arXiv},
       eprint = {2004.08804},
 primaryClass = {astro-ph.SR},
       adsurl = {https://ui.adsabs.harvard.edu/abs/2020A&A...640A..82T}
}

@ARTICLE{Taboneetal2017_SO_SO2,
       author = {{Tabone}, B. and {Cabrit}, S. and {Bianchi}, E. and {Ferreira}, J. and {Pineau des For{\^e}ts}, G. and {Codella}, C. and {Gusdorf}, A. and {Gueth}, F. and {Podio}, L. and {Chapillon}, E.},
        title = "{ALMA discovery of a rotating SO/SO$_{2}$ flow in HH212. A possible MHD disk wind?}",
      journal = {\aap},
         year = 2017,
        month = nov,
       volume = {607},
          eid = {L6},
        pages = {L6},
          doi = {10.1051/0004-6361/201731691},
archivePrefix = {arXiv},
       eprint = {1710.01401},
 primaryClass = {astro-ph.SR},
       adsurl = {https://ui.adsabs.harvard.edu/abs/2017A&A...607L...6T}
}

@ARTICLE{Bianchietal2017_HH212methanol,
       author = {{Bianchi}, E. and {Codella}, C. and {Ceccarelli}, C. and {Taquet}, V. and {Cabrit}, S. and {Bacciotti}, F. and {Bachiller}, R. and {Chapillon}, E. and {Gueth}, F. and {Gusdorf}, A. and {Lefloch}, B. and {Leurini}, S. and {Podio}, L. and {Rygl}, K.~L.~J. and {Tabone}, B. and {Tafalla}, M.},
        title = "{Deuterated methanol on a solar system scale around the HH212 protostar}",
      journal = {\aap},
         year = 2017,
        month = oct,
       volume = {606},
          eid = {L7},
        pages = {L7},
          doi = {10.1051/0004-6361/201731404},
archivePrefix = {arXiv},
       eprint = {1709.04726},
 primaryClass = {astro-ph.SR},
       adsurl = {https://ui.adsabs.harvard.edu/abs/2017A&A...606L...7B}
}

@ARTICLE{Codellaetal2014_HH212,
       author = {{Codella}, C. and {Cabrit}, S. and {Gueth}, F. and {Podio}, L. and {Leurini}, S. and {Bachiller}, R. and {Gusdorf}, A. and {Lefloch}, B. and {Nisini}, B. and {Tafalla}, M. and {Yvart}, W.},
        title = "{The ALMA view of the protostellar system HH212. The wind, the cavity, and the disk}",
      journal = {\aap},
         year = 2014,
        month = aug,
       volume = {568},
          eid = {L5},
        pages = {L5},
          doi = {10.1051/0004-6361/201424103},
archivePrefix = {arXiv},
       eprint = {1407.6229},
 primaryClass = {astro-ph.SR},
       adsurl = {https://ui.adsabs.harvard.edu/abs/2014A&A...568L...5C}
}

@ARTICLE{Podio2015_SO,
       author = {{Podio}, L. and {Codella}, C. and {Gueth}, F. and {Cabrit}, S. and {Bachiller}, R. and {Gusdorf}, A. and {Lee}, C. -F. and {Lefloch}, B. and {Leurini}, S. and {Nisini}, B. and {Tafalla}, M.},
        title = "{The jet and the disk of the HH 212 low-mass protostar imaged by ALMA: SO and SO$_{2}$ emission}",
      journal = {\aap},
         year = 2015,
        month = sep,
       volume = {581},
          eid = {A85},
        pages = {A85},
          doi = {10.1051/0004-6361/201525778},
archivePrefix = {arXiv},
       eprint = {1505.05919},
 primaryClass = {astro-ph.GA},
       adsurl = {https://ui.adsabs.harvard.edu/abs/2015A&A...581A..85P}
}

@ARTICLE{Codellaetal2018_HDO,
       author = {{Codella}, C. and {Bianchi}, E. and {Tabone}, B. and {Lee}, C. -F. and {Cabrit}, S. and {Ceccarelli}, C. and {Podio}, L. and {Bacciotti}, F. and {Bachiller}, R. and {Chapillon}, E. and {Gueth}, F. and {Gusdorf}, A. and {Lefloch}, B. and {Leurini}, S. and {Pineau des For{\^e}ts}, G. and {Rygl}, K.~L.~J. and {Tafalla}, M.},
        title = "{Water and interstellar complex organics associated with the HH 212 protostellar disc. On disc atmospheres, disc winds, and accretion shocks}",
      journal = {\aap},
         year = 2018,
        month = sep,
       volume = {617},
          eid = {A10},
        pages = {A10},
          doi = {10.1051/0004-6361/201832765},
archivePrefix = {arXiv},
       eprint = {1806.07967},
 primaryClass = {astro-ph.SR},
       adsurl = {https://ui.adsabs.harvard.edu/abs/2018A&A...617A..10C}
}

@ARTICLE{Codellaetal2016_HDO,
       author = {{Codella}, C. and {Ceccarelli}, C. and {Cabrit}, S. and {Gueth}, F. and {Podio}, L. and {Bachiller}, R. and {Fontani}, F. and {Gusdorf}, A. and {Lefloch}, B. and {Leurini}, S. and {Tafalla}, M.},
        title = "{Water and acetaldehyde in HH212: The first hot corino in Orion}",
      journal = {\aap},
         year = 2016,
        month = feb,
       volume = {586},
          eid = {L3},
        pages = {L3},
          doi = {10.1051/0004-6361/201527424},
archivePrefix = {arXiv},
       eprint = {1601.00539},
 primaryClass = {astro-ph.SR},
       adsurl = {https://ui.adsabs.harvard.edu/abs/2016A&A...586L...3C}
}

@ARTICLE{Cabritetal2012,
       author = {{Cabrit}, S. and {Codella}, C. and {Gueth}, F. and {Gusdorf}, A.},
        title = "{High SiO abundance in the HH212 protostellar jet}",
      journal = {\aap},
         year = 2012,
        month = dec,
       volume = {548},
          eid = {L2},
        pages = {L2},
          doi = {10.1051/0004-6361/201219784},
archivePrefix = {arXiv},
       eprint = {1211.1258},
 primaryClass = {astro-ph.GA},
       adsurl = {https://ui.adsabs.harvard.edu/abs/2012A&A...548L...2C}
}

@ARTICLE{reipurthetal2019,
       author = {{Reipurth}, Bo and {Davis}, C.~J. and {Bally}, John and {Raga}, A.~C. and {Bowler}, B.~P. and {Geballe}, T.~R. and {Aspin}, Colin and {Chiang}, Hsin-Fang},
        title = "{The Giant Herbig-Haro Flow HH 212 and Associated Star Formation}",
      journal = {\aj},
         year = 2019,
        month = sep,
       volume = {158},
       number = {3},
          eid = {107},
        pages = {107},
          doi = {10.3847/1538-3881/ab2d25},
archivePrefix = {arXiv},
       eprint = {1911.05877},
 primaryClass = {astro-ph.SR},
       adsurl = {https://ui.adsabs.harvard.edu/abs/2019AJ....158..107R}
}

@ARTICLE{Zinneckeretal1998,
       author = {{Zinnecker}, Hans and {McCaughrean}, Mark J. and {Rayner}, John T.},
        title = "{A symmetrically pulsed jet of gas from an invisible protostar in Orion}",
      journal = {\nat},
         year = 1998,
        month = aug,
       volume = {394},
       number = {6696},
        pages = {862-865},
          doi = {10.1038/29716},
       adsurl = {https://ui.adsabs.harvard.edu/abs/1998Natur.394..862Z}
}

@ARTICLE{Legouellec2020,
       author = {{Le Gouellec}, V.~J.~M. and {Maury}, A.~J. and {Guillet}, V. and {Hull}, C.~L.~H. and {Girart}, J.~M. and {Verliat}, A. and {Mignon-Risse}, R. and {Valdivia}, V. and {Hennebelle}, P. and {Gonz{\'a}lez}, M. and {Louvet}, F.},
        title = "{A statistical analysis of dust polarization properties in ALMA observations of Class 0 protostellar cores}",
      journal = {\aap},
         year = 2020,
        month = dec,
       volume = {644},
          eid = {A11},
        pages = {A11},
          doi = {10.1051/0004-6361/202038404},
archivePrefix = {arXiv},
       eprint = {2009.07186},
 primaryClass = {astro-ph.GA},
       adsurl = {https://ui.adsabs.harvard.edu/abs/2020A&A...644A..11L}
}

@ARTICLE{maury2018,
       author = {{Maury}, A.~J. and {Girart}, J.~M. and {Zhang}, Q. and {Hennebelle}, P. and {Keto}, E. and {Rao}, R. and {Lai}, S. -P. and {Ohashi}, N. and {Galametz}, M.},
        title = "{Magnetically regulated collapse in the B335 protostar? I. ALMA observations of the polarized dust emission}",
      journal = {\mnras},
         year = 2018,
        month = jun,
       volume = {477},
       number = {2},
        pages = {2760-2765},
          doi = {10.1093/mnras/sty574},
archivePrefix = {arXiv},
       eprint = {1803.00028},
 primaryClass = {astro-ph.GA},
       adsurl = {https://ui.adsabs.harvard.edu/abs/2018MNRAS.477.2760M}
}

@ARTICLE{maury2012,
       author = {{Maury}, A. and {Ohashi}, N. and {Andr{\'e}}, P.},
        title = "{Subarcsecond SMA observations of the prototype Class 0 object VLA1623 at 1.3 mm: a single protostar with a structured outflow cavity?}",
      journal = {\aap},
         year = 2012,
        month = mar,
       volume = {539},
          eid = {A130},
        pages = {A130},
          doi = {10.1051/0004-6361/201118180},
archivePrefix = {arXiv},
       eprint = {1201.6395},
 primaryClass = {astro-ph.GA},
       adsurl = {https://ui.adsabs.harvard.edu/abs/2012A&A...539A.130M}
}

@ARTICLE{Kounkel17,
       author = {{Kounkel}, Marina and {Covey}, Kevin and {Su{\'a}rez}, Genaro and {Rom{\'a}n-Z{\'u}{\~n}iga}, Carlos and {Hernandez}, Jesus and {Stassun}, Keivan and {Jaehnig}, Karl O. and {Feigelson}, Eric D. and {Pe{\~n}a Ram{\'\i}rez}, Karla and {Roman-Lopes}, Alexandre and {Da Rio}, Nicola and {Stringfellow}, Guy S. and {Kim}, J. Serena and {Borissova}, Jura and {Fern{\'a}ndez-Trincado}, Jos{\'e} G. and {Burgasser}, Adam and {Garc{\'\i}a-Hern{\'a}ndez}, D.~A. and {Zamora}, Olga and {Pan}, Kaike and {Nitschelm}, Christian},
        title = "{The APOGEE-2 Survey of the Orion Star-forming Complex. II. Six-dimensional Structure}",
      journal = {\aj},
         year = 2018,
        month = sep,
       volume = {156},
       number = {3},
          eid = {84},
        pages = {84},
          doi = {10.3847/1538-3881/aad1f1},
archivePrefix = {arXiv},
       eprint = {1805.04649},
 primaryClass = {astro-ph.SR},
       adsurl = {https://ui.adsabs.harvard.edu/abs/2018AJ....156...84K}
}

@ARTICLE{Motte01,
       author = {{Motte}, F. and {Andr{\'e}}, P.},
        title = "{The circumstellar environment of low-mass protostars: A millimeter continuum mapping survey}",
      journal = {\aap},
         year = 2001,
        month = jan,
       volume = {365},
        pages = {440-464},
          doi = {10.1051/0004-6361:20000072},
       adsurl = {https://ui.adsabs.harvard.edu/abs/2001A&A...365..440M}
}

@ARTICLE{Galametzetal2019,
       author = {{Galametz}, M. and {Maury}, A.~J. and {Valdivia}, V. and {Testi}, L. and {Belloche}, A. and {Andr{\'e}}, Ph.},
        title = "{Low dust emissivities and radial variations in the envelopes of Class 0 protostars: possible signature of early grain growth}",
      journal = {\aap},
         year = 2019,
        month = dec,
       volume = {632},
          eid = {A5},
        pages = {A5},
          doi = {10.1051/0004-6361/201936342},
archivePrefix = {arXiv},
       eprint = {1910.04652},
 primaryClass = {astro-ph.GA},
       adsurl = {https://ui.adsabs.harvard.edu/abs/2019A&A...632A...5G}
}

@ARTICLE{Tobinetal2020,
       author = {{Tobin}, John J. and {Sheehan}, Patrick D. and {Megeath}, S. Thomas and {D{\'\i}az-Rodr{\'\i}guez}, Ana Karla and {Offner}, Stella S.~R. and {Murillo}, Nadia M. and {van 't Hoff}, Merel L.~R. and {van Dishoeck}, Ewine F. and {Osorio}, Mayra and {Anglada}, Guillem and {Furlan}, Elise and {Stutz}, Amelia M. and {Reynolds}, Nickalas and {Karnath}, Nicole and {Fischer}, William J. and {Persson}, Magnus and {Looney}, Leslie W. and {Li}, Zhi-Yun and {Stephens}, Ian and {Chandler}, Claire J. and {Cox}, Erin and {Dunham}, Michael M. and {Tychoniec}, {\L}ukasz and {Kama}, Mihkel and {Kratter}, Kaitlin and {Kounkel}, Marina and {Mazur}, Brian and {Maud}, Luke and {Patel}, Lisa and {Perez}, Laura and {Sadavoy}, Sarah I. and {Segura-Cox}, Dominique and {Sharma}, Rajeeb and {Stephenson}, Brian and {Watson}, Dan M. and {Wyrowski}, Friedrich},
        title = "{The VLA/ALMA Nascent Disk and Multiplicity (VANDAM) Survey of Orion Protostars. II. A Statistical Characterization of Class 0 and Class I Protostellar Disks}",
      journal = {\apj},
         year = 2020,
        month = feb,
       volume = {890},
       number = {2},
          eid = {130},
        pages = {130},
          doi = {10.3847/1538-4357/ab6f64},
archivePrefix = {arXiv},
       eprint = {2001.04468},
 primaryClass = {astro-ph.GA},
       adsurl = {https://ui.adsabs.harvard.edu/abs/2020ApJ...890..130T}
}

@INPROCEEDINGS{Bouvier14,
       author = {{Bouvier}, J. and {Matt}, S.~P. and {Mohanty}, S. and {Scholz}, A. and {Stassun}, K.~G. and {Zanni}, C.},
        title = "{Angular Momentum Evolution of Young Low-Mass Stars and Brown Dwarfs: Observations and Theory}",
    booktitle = {Protostars and Planets VI},
         year = 2014,
       editor = {{Beuther}, Henrik and {Klessen}, Ralf S. and {Dullemond}, Cornelis P. and {Henning}, Thomas},
        month = jan,
        pages = {433},
          doi = {10.2458/azu\_uapress\_9780816531240-ch019},
archivePrefix = {arXiv},
       eprint = {1309.7851},
 primaryClass = {astro-ph.SR},
       adsurl = {https://ui.adsabs.harvard.edu/abs/2014prpl.conf..433B}
}

@INPROCEEDINGS{Franketal2014_PPVI,
       author = {{Frank}, A. and {Ray}, T.~P. and {Cabrit}, S. and {Hartigan}, P. and {Arce}, H.~G. and {Bacciotti}, F. and {Bally}, J. and {Benisty}, M. and {Eisl{\"o}ffel}, J. and {G{\"u}del}, M. and {Lebedev}, S. and {Nisini}, B. and {Raga}, A.},
        title = "{Jets and Outflows from Star to Cloud: Observations Confront Theory}",
    booktitle = {Protostars and Planets VI},
         year = 2014,
       editor = {{Beuther}, Henrik and {Klessen}, Ralf S. and {Dullemond}, Cornelis P. and {Henning}, Thomas},
        month = jan,
        pages = {451},
          doi = {10.2458/azu\_uapress\_9780816531240-ch020},
archivePrefix = {arXiv},
       eprint = {1402.3553},
 primaryClass = {astro-ph.SR},
       adsurl = {https://ui.adsabs.harvard.edu/abs/2014prpl.conf..451F}
}

@INPROCEEDINGS{Testietal2014_PPVI,
       author = {{Testi}, L. and {Birnstiel}, T. and {Ricci}, L. and {Andrews}, S. and {Blum}, J. and {Carpenter}, J. and {Dominik}, C. and {Isella}, A. and {Natta}, A. and {Williams}, J.~P. and {Wilner}, D.~J.},
        title = "{Dust Evolution in Protoplanetary Disks}",
    booktitle = {Protostars and Planets VI},
         year = 2014,
       editor = {{Beuther}, Henrik and {Klessen}, Ralf S. and {Dullemond}, Cornelis P. and {Henning}, Thomas},
        month = jan,
        pages = {339},
          doi = {10.2458/azu\_uapress\_9780816531240-ch015},
archivePrefix = {arXiv},
       eprint = {1402.1354},
 primaryClass = {astro-ph.SR},
       adsurl = {https://ui.adsabs.harvard.edu/abs/2014prpl.conf..339T}
}

@INPROCEEDINGS{Pascucci2023,
       author = {{Pascucci}, I. and {Cabrit}, S. and {Edwards}, S. and {Gorti}, U. and {Gressel}, O. and {Suzuki}, T.~K.},
        title = "{The Role of Disk Winds in the Evolution and Dispersal of Protoplanetary Disks}",
    booktitle = {Protostars and Planets VII},
         year = 2023,
       editor = {{Inutsuka}, S. and {Aikawa}, Y. and {Muto}, T. and {Tomida}, K. and {Tamura}, M.},
       series = {Astronomical Society of the Pacific Conference Series},
       volume = {534},
        month = jul,
        pages = {567},
          doi = {10.48550/arXiv.2203.10068},
archivePrefix = {arXiv},
       eprint = {2203.10068},
 primaryClass = {astro-ph.EP},
       adsurl = {https://ui.adsabs.harvard.edu/abs/2023ASPC..534..567P}
}

@ARTICLE{Claussen1998,
       author = {{Claussen}, Mark J. and {Marvel}, Kevin B. and {Wootten}, Alwyn and {Wilking}, Bruce A.},
        title = "{Distribution and Motion of the Water Masers near IRAS 05413-0104}",
      journal = {\apjl},
         year = 1998,
        month = nov,
       volume = {507},
       number = {1},
        pages = {L79-L82},
          doi = {10.1086/311669},
       adsurl = {https://ui.adsabs.harvard.edu/abs/1998ApJ...507L..79C}
}

@ARTICLE{devalonetal2020_dgtaub,
       author = {{de Valon}, A. and {Dougados}, C. and {Cabrit}, S. and {Louvet}, F. and {Zapata}, L.~A. and {Mardones}, D.},
        title = "{ALMA reveals a large structured disk and nested rotating outflows in DG Tauri B}",
      journal = {\aap},
         year = 2020,
        month = feb,
       volume = {634},
          eid = {L12},
        pages = {L12},
          doi = {10.1051/0004-6361/201936950},
archivePrefix = {arXiv},
       eprint = {2001.09776},
 primaryClass = {astro-ph.SR},
       adsurl = {https://ui.adsabs.harvard.edu/abs/2020A&A...634L..12D}
}

@article{Bjerkeli19,
   title={Kinematics around the B335 protostar down to au scales},
   volume={631},
   ISSN={1432-0746},
   url={http://dx.doi.org/10.1051/0004-6361/201935948},
   DOI={10.1051/0004-6361/201935948},
   journal={\aap},
   publisher={EDP Sciences},
   author={Bjerkeli, Per and Ramsey, Jon P. and Harsono, Daniel and Calcutt, Hannah and Kristensen, Lars E. and van der Wiel, Matthijs H. D. and Jørgensen, Jes K. and Muller, Sébastien and Persson, Magnus V.},
   year={2019},
   month={Oct},
   pages={A64}
}

@Article{bjerkeli2016,
  Title                    = {{Resolved images of a protostellar outflow driven by an extended disk wind}},
  Author                   = {{Bjerkeli}, P. and {van der Wiel}, M.~H.~D. and {Harsono}, D. and 
 {Ramsey}, J.~P. and {J{\o}rgensen}, J.~K.},
  Journal                  = {\nat},
  Year                     = {2016},
  Month                    = dec,
  Pages                    = {406-409},
  Volume                   = {540},
  Adsurl                   = {http://adsabs.harvard.edu/abs/2016Natur.540..406B},
  Archiveprefix            = {arXiv},
  Doi                      = {10.1038/nature20600},
  Eprint                   = {1612.05148},
  Primaryclass             = {astro-ph.SR}
}

@ARTICLE{Snelletal80,
   author = {{Snell}, R.~L. and {Loren}, R.~B. and {Plambeck}, R.~L.},
    title = "{Observations of CO in L1551 - Evidence for stellar wind driven shocks}",
  journal = {\apjl},
     year = 1980,
    month = jul,
   volume = 239,
    pages = {L17-L22},
      doi = {10.1086/183283},
   adsurl = {http://adsabs.harvard.edu/abs/1980ApJ...239L..17S}
}

@ARTICLE{Hartiganetal2005,
   author = {{Hartigan}, P. and {Heathcote}, S. and {Morse}, J.~A. and {Reipurth}, B. and 
        {Bally}, J.},
    title = "{Proper Motions of the HH 47 Jet Observed with the Hubble Space Telescope}",
  journal = {\aj},
   eprint = {astro-ph/0507526},
     year = 2005,
    month = nov,
   volume = 130,
    pages = {2197-2205},
      doi = {10.1086/491673},
   adsurl = {http://adsabs.harvard.edu/abs/2005AJ....130.2197H}
}

@ARTICLE{Podioetal2006,
   author = {{Podio}, L. and {Bacciotti}, F. and {Nisini}, B. and {Eisl{\"o}ffel}, J. and 
	{Massi}, F. and {Giannini}, T. and {Ray}, T.~P.},
    title = "{Recipes for stellar jets: results of combined optical/infrared diagnostics}",
  journal = {\aap},
   eprint = {astro-ph/0606280},
     year = 2006,
    month = sep,
   volume = 456,
    pages = {189-204},
      doi = {10.1051/0004-6361:20054156},
   adsurl = {http://adsabs.harvard.edu/abs/2006A%26A...456..189P}
}

@ARTICLE{Plunkettetal2015,
       author = {{Plunkett}, Adele L. and {Arce}, H{\'e}ctor G. and {Mardones}, Diego and {van Dokkum}, Pieter and {Dunham}, Michael M. and {Fern{\'a}ndez-L{\'o}pez}, Manuel and {Gallardo}, Jos{\'e} and {Corder}, Stuartt A.},
        title = "{Episodic molecular outflow in the very young protostellar cluster Serpens South}",
      journal = {\nat},
         year = 2015,
        month = nov,
       volume = {527},
       number = {7576},
        pages = {70-73},
          doi = {10.1038/nature15702},
archivePrefix = {arXiv},
       eprint = {1511.01100},
 primaryClass = {astro-ph.SR},
       adsurl = {https://ui.adsabs.harvard.edu/abs/2015Natur.527...70P}
}

@ARTICLE{Kristensenetal2017,
       author = {{Kristensen}, L.~E. and {Gusdorf}, A. and {Mottram}, J.~C. and {Karska}, A. and {Visser}, R. and {Wiesemeyer}, H. and {G{\"u}sten}, R. and {Simon}, R.},
        title = "{Oxygen budget in low-mass protostars: the NGC 1333-IRAS4A R1 shock observed in [O I] at 63 {\ensuremath{\mu}}m with SOFIA-GREAT}",
      journal = {\aap},
         year = 2017,
        month = may,
       volume = {601},
          eid = {L4},
        pages = {L4},
          doi = {10.1051/0004-6361/201630310},
archivePrefix = {arXiv},
       eprint = {1704.07593},
 primaryClass = {astro-ph.SR},
       adsurl = {https://ui.adsabs.harvard.edu/abs/2017A&A...601L...4K}
}

@ARTICLE{Bjerkelietal2012,
       author = {{Bjerkeli}, P. and {Liseau}, R. and {Larsson}, B. and {Rydbeck}, G. and
         {Nisini}, B. and {Tafalla}, M. and {Antoniucci}, S. and
         {Benedettini}, M. and {Bergman}, P. and {Cabrit}, S.},
        title = "{H$_{2}$O line mapping at high spatial and spectral resolution. Herschel observations of the VLA 1623 outflow}",
      journal = {\aap},
         year = "2012",
        month = "Oct",
       volume = {546},
          eid = {A29},
        pages = {A29},
          doi = {10.1051/0004-6361/201219776},
archivePrefix = {arXiv},
       eprint = {1209.0294},
 primaryClass = {astro-ph.GA},
       adsurl = {https://ui.adsabs.harvard.edu/abs/2012A&A...546A..29B}
}

@ARTICLE{Liuetal2016,
       author = {{Liu}, Chun-Fan and {Shang}, Hsien and {Herczeg}, Gregory J. and {Walter}, Frederick M.},
        title = "{The [Ne III] Jet of DG Tau and its Ionization Scenarios}",
      journal = {\apj},
         year = 2016,
        month = dec,
       volume = {832},
       number = {2},
          eid = {153},
        pages = {153},
          doi = {10.3847/0004-637X/832/2/153},
archivePrefix = {arXiv},
       eprint = {1611.01953},
 primaryClass = {astro-ph.SR},
       adsurl = {https://ui.adsabs.harvard.edu/abs/2016ApJ...832..153L}
}

@ARTICLE{GresselRamseyetal2020_diskwinds,
       author = {{Gressel}, Oliver and {Ramsey}, Jon P. and {Brinch}, Christian and {Nelson}, Richard P. and {Turner}, Neal J. and {Bruderer}, Simon},
        title = "{Global Hydromagnetic Simulations of Protoplanetary Disks with Stellar Irradiation and Simplified Thermochemistry}",
      journal = {\apj},
         year = 2020,
        month = jun,
       volume = {896},
       number = {2},
          eid = {126},
        pages = {126},
          doi = {10.3847/1538-4357/ab91b7},
archivePrefix = {arXiv},
       eprint = {2005.03431},
 primaryClass = {astro-ph.EP},
       adsurl = {https://ui.adsabs.harvard.edu/abs/2020ApJ...896..126G}
}

@ARTICLE{Gresseletal2015_diskwinds,
       author = {{Gressel}, Oliver and {Turner}, Neal J. and {Nelson}, Richard P. and {McNally}, Colin P.},
        title = "{Global Simulations of Protoplanetary Disks With Ohmic Resistivity and Ambipolar Diffusion}",
      journal = {\apj},
         year = 2015,
        month = mar,
       volume = {801},
       number = {2},
          eid = {84},
        pages = {84},
          doi = {10.1088/0004-637X/801/2/84},
archivePrefix = {arXiv},
       eprint = {1501.05431},
 primaryClass = {astro-ph.EP},
       adsurl = {https://ui.adsabs.harvard.edu/abs/2015ApJ...801...84G}
}

@ARTICLE{Sheikhnezamietal2012_diskwinds,
       author = {{Sheikhnezami}, Somayeh and {Fendt}, Christian and {Porth}, Oliver and {Vaidya}, Bhargav and {Ghanbari}, Jamshid},
        title = "{Bipolar Jets Launched from Magnetically Diffusive Accretion Disks. I. Ejection Efficiency versus Field Strength and Diffusivity}",
      journal = {\apj},
         year = 2012,
        month = sep,
       volume = {757},
       number = {1},
          eid = {65},
        pages = {65},
          doi = {10.1088/0004-637X/757/1/65},
archivePrefix = {arXiv},
       eprint = {1207.6086},
 primaryClass = {astro-ph.HE},
       adsurl = {https://ui.adsabs.harvard.edu/abs/2012ApJ...757...65S}
}

@ARTICLE{RamseyClarke2019_diskwinds,
       author = {{Ramsey}, Jon P. and {Clarke}, David A.},
        title = "{MHD simulations of the formation and propagation of protostellar jets to observational length-scales}",
      journal = {\mnras},
         year = 2019,
        month = apr,
       volume = {484},
       number = {2},
        pages = {2364-2387},
          doi = {10.1093/mnras/stz116},
archivePrefix = {arXiv},
       eprint = {1901.02845},
 primaryClass = {astro-ph.SR},
       adsurl = {https://ui.adsabs.harvard.edu/abs/2019MNRAS.484.2364R}
}

@ARTICLE{PudritzNorman1983_diskwinds,
       author = {{Pudritz}, R.~E. and {Norman}, C.~A.},
        title = "{Centrifugally driven winds from contracting molecular disks}",
      journal = {\apj},
         year = 1983,
        month = nov,
       volume = {274},
        pages = {677-697},
          doi = {10.1086/161481},
       adsurl = {https://ui.adsabs.harvard.edu/abs/1983ApJ...274..677P}
}

@ARTICLE{BlandfordPayne1982,
       author = {{Blandford}, R.~D. and {Payne}, D.~G.},
        title = "{Hydromagnetic flows from accretion disks and the production of radio jets.}",
      journal = {\mnras},
         year = 1982,
        month = jun,
       volume = {199},
        pages = {883-903},
          doi = {10.1093/mnras/199.4.883},
       adsurl = {https://ui.adsabs.harvard.edu/abs/1982MNRAS.199..883B}
}

@ARTICLE{CASA2022,
       author = {{CASA Team} and {Bean}, Ben and {Bhatnagar}, Sanjay and {Castro}, Sandra and {Donovan Meyer}, Jennifer and {Emonts}, Bjorn and {Garcia}, Enrique and {Garwood}, Robert and {Golap}, Kumar and {Gonzalez Villalba}, Justo and {Harris}, Pamela and {Hayashi}, Yohei and {Hoskins}, Josh and {Hsieh}, Mingyu and {Jagannathan}, Preshanth and {Kawasaki}, Wataru and {Keimpema}, Aard and {Kettenis}, Mark and {Lopez}, Jorge and {Marvil}, Joshua and {Masters}, Joseph and {McNichols}, Andrew and {Mehringer}, David and {Miel}, Renaud and {Moellenbrock}, George and {Montesino}, Federico and {Nakazato}, Takeshi and {Ott}, Juergen and {Petry}, Dirk and {Pokorny}, Martin and {Raba}, Ryan and {Rau}, Urvashi and {Schiebel}, Darrell and {Schweighart}, Neal and {Sekhar}, Srikrishna and {Shimada}, Kazuhiko and {Small}, Des and {Steeb}, Jan-Willem and {Sugimoto}, Kanako and {Suoranta}, Ville and {Tsutsumi}, Takahiro and {van Bemmel}, Ilse M. and {Verkouter}, Marjolein and {Wells}, Akeem and {Xiong}, Wei and {Szomoru}, Arpad and {Griffith}, Morgan and {Glendenning}, Brian and {Kern}, Jeff},
        title = "{CASA, the Common Astronomy Software Applications for Radio Astronomy}",
      journal = {\pasp},
         year = 2022,
        month = nov,
       volume = {134},
       number = {1041},
          eid = {114501},
        pages = {114501},
          doi = {10.1088/1538-3873/ac9642},
archivePrefix = {arXiv},
       eprint = {2210.02276},
 primaryClass = {astro-ph.IM},
       adsurl = {https://ui.adsabs.harvard.edu/abs/2022PASP..134k4501C}
}

@INPROCEEDINGS{McMullingetal2007_CASA,
       author = {{McMullin}, J.~P. and {Waters}, B. and {Schiebel}, D. and {Young}, W. and {Golap}, K.},
        title = "{CASA Architecture and Applications}",
    booktitle = {Astronomical Data Analysis Software and Systems XVI},
         year = 2007,
       editor = {{Shaw}, R.~A. and {Hill}, F. and {Bell}, D.~J.},
       series = {Astronomical Society of the Pacific Conference Series},
       volume = {376},
        month = oct,
        pages = {127},
       adsurl = {https://ui.adsabs.harvard.edu/abs/2007ASPC..376..127M}
}

@Article{Hunter2007_matplotlib,
  Author    = {Hunter, J. D.},
  Title     = {Matplotlib: A 2D graphics environment},
  Journal   = {Computing in Science \& Engineering},
  Volume    = {9},
  Number    = {3},
  Pages     = {90--95},
  publisher = {IEEE COMPUTER SOC},
  doi       = {10.1109/MCSE.2007.55},
  year      = 2007
}

@ARTICLE{Astropy2013_I,
       author = {{Astropy Collaboration} and {Robitaille}, Thomas P. and {Tollerud}, Erik J. and {Greenfield}, Perry and {Droettboom}, Michael and {Bray}, Erik and {Aldcroft}, Tom and {Davis}, Matt and {Ginsburg}, Adam and {Price-Whelan}, Adrian M. and {Kerzendorf}, Wolfgang E. and {Conley}, Alexander and {Crighton}, Neil and {Barbary}, Kyle and {Muna}, Demitri and {Ferguson}, Henry and {Grollier}, Fr{\'e}d{\'e}ric and {Parikh}, Madhura M. and {Nair}, Prasanth H. and {Unther}, Hans M. and {Deil}, Christoph and {Woillez}, Julien and {Conseil}, Simon and {Kramer}, Roban and {Turner}, James E.~H. and {Singer}, Leo and {Fox}, Ryan and {Weaver}, Benjamin A. and {Zabalza}, Victor and {Edwards}, Zachary I. and {Azalee Bostroem}, K. and {Burke}, D.~J. and {Casey}, Andrew R. and {Crawford}, Steven M. and {Dencheva}, Nadia and {Ely}, Justin and {Jenness}, Tim and {Labrie}, Kathleen and {Lim}, Pey Lian and {Pierfederici}, Francesco and {Pontzen}, Andrew and {Ptak}, Andy and {Refsdal}, Brian and {Servillat}, Mathieu and {Streicher}, Ole},
        title = "{Astropy: A community Python package for astronomy}",
      journal = {\aap},
         year = 2013,
        month = oct,
       volume = {558},
          eid = {A33},
        pages = {A33},
          doi = {10.1051/0004-6361/201322068},
archivePrefix = {arXiv},
       eprint = {1307.6212},
 primaryClass = {astro-ph.IM},
       adsurl = {https://ui.adsabs.harvard.edu/abs/2013A&A...558A..33A}
}

@ARTICLE{Astropy2018_II,
       author = {{Astropy Collaboration} and {Price-Whelan}, A.~M. and {Sip{\H{o}}cz}, B.~M. and {G{\"u}nther}, H.~M. and {Lim}, P.~L. and {Crawford}, S.~M. and {Conseil}, S. and {Shupe}, D.~L. and {Craig}, M.~W. and {Dencheva}, N. and {Ginsburg}, A. and {VanderPlas}, J.~T. and {Bradley}, L.~D. and {P{\'e}rez-Su{\'a}rez}, D. and {de Val-Borro}, M. and {Aldcroft}, T.~L. and {Cruz}, K.~L. and {Robitaille}, T.~P. and {Tollerud}, E.~J. and {Ardelean}, C. and {Babej}, T. and {Bach}, Y.~P. and {Bachetti}, M. and {Bakanov}, A.~V. and {Bamford}, S.~P. and {Barentsen}, G. and {Barmby}, P. and {Baumbach}, A. and {Berry}, K.~L. and {Biscani}, F. and {Boquien}, M. and {Bostroem}, K.~A. and {Bouma}, L.~G. and {Brammer}, G.~B. and {Bray}, E.~M. and {Breytenbach}, H. and {Buddelmeijer}, H. and {Burke}, D.~J. and {Calderone}, G. and {Cano Rodr{\'\i}guez}, J.~L. and {Cara}, M. and {Cardoso}, J.~V.~M. and {Cheedella}, S. and {Copin}, Y. and {Corrales}, L. and {Crichton}, D. and {D'Avella}, D. and {Deil}, C. and {Depagne}, {\'E}. and {Dietrich}, J.~P. and {Donath}, A. and {Droettboom}, M. and {Earl}, N. and {Erben}, T. and {Fabbro}, S. and {Ferreira}, L.~A. and {Finethy}, T. and {Fox}, R.~T. and {Garrison}, L.~H. and {Gibbons}, S.~L.~J. and {Goldstein}, D.~A. and {Gommers}, R. and {Greco}, J.~P. and {Greenfield}, P. and {Groener}, A.~M. and {Grollier}, F. and {Hagen}, A. and {Hirst}, P. and {Homeier}, D. and {Horton}, A.~J. and {Hosseinzadeh}, G. and {Hu}, L. and {Hunkeler}, J.~S. and {Ivezi{\'c}}, {\v{Z}}. and {Jain}, A. and {Jenness}, T. and {Kanarek}, G. and {Kendrew}, S. and {Kern}, N.~S. and {Kerzendorf}, W.~E. and {Khvalko}, A. and {King}, J. and {Kirkby}, D. and {Kulkarni}, A.~M. and {Kumar}, A. and {Lee}, A. and {Lenz}, D. and {Littlefair}, S.~P. and {Ma}, Z. and {Macleod}, D.~M. and {Mastropietro}, M. and {McCully}, C. and {Montagnac}, S. and {Morris}, B.~M. and {Mueller}, M. and {Mumford}, S.~J. and {Muna}, D. and {Murphy}, N.~A. and {Nelson}, S. and {Nguyen}, G.~H. and {Ninan}, J.~P. and {N{\"o}the}, M. and {Ogaz}, S. and {Oh}, S. and {Parejko}, J.~K. and {Parley}, N. and {Pascual}, S. and {Patil}, R. and {Patil}, A.~A. and {Plunkett}, A.~L. and {Prochaska}, J.~X. and {Rastogi}, T. and {Reddy Janga}, V. and {Sabater}, J. and {Sakurikar}, P. and {Seifert}, M. and {Sherbert}, L.~E. and {Sherwood-Taylor}, H. and {Shih}, A.~Y. and {Sick}, J. and {Silbiger}, M.~T. and {Singanamalla}, S. and {Singer}, L.~P. and {Sladen}, P.~H. and {Sooley}, K.~A. and {Sornarajah}, S. and {Streicher}, O. and {Teuben}, P. and {Thomas}, S.~W. and {Tremblay}, G.~R. and {Turner}, J.~E.~H. and {Terr{\'o}n}, V. and {van Kerkwijk}, M.~H. and {de la Vega}, A. and {Watkins}, L.~L. and {Weaver}, B.~A. and {Whitmore}, J.~B. and {Woillez}, J. and {Zabalza}, V. and {Astropy Contributors}},
        title = "{The Astropy Project: Building an Open-science Project and Status of the v2.0 Core Package}",
      journal = {\aj},
         year = 2018,
        month = sep,
       volume = {156},
       number = {3},
          eid = {123},
        pages = {123},
          doi = {10.3847/1538-3881/aabc4f},
archivePrefix = {arXiv},
       eprint = {1801.02634},
 primaryClass = {astro-ph.IM},
       adsurl = {https://ui.adsabs.harvard.edu/abs/2018AJ....156..123A}
}

@conference{Kluyveretal2016_jupyter,
  Title = {Jupyter Notebooks -- a publishing format for reproducible computational workflows},
  Author = {Thomas Kluyver and Benjamin Ragan-Kelley and Fernando P{\'e}rez and Brian Granger and Matthias Bussonnier and Jonathan Frederic and Kyle Kelley and Jessica Hamrick and Jason Grout and Sylvain Corlay and Paul Ivanov and Dami{\'a}n Avila and Safia Abdalla and Carol Willing},
  Booktitle = {Positioning and Power in Academic Publishing: Players, Agents and Agendas},
  Editor = {F. Loizides and B. Schmidt},
  Organization = {IOS Press},
  Pages = {87 - 90},
  Year = {2016}
}

@misc{pandas,
    author       = {The pandas development team},
    title        = {pandas-dev/pandas: Pandas},
    month        = feb,
    year         = 2020,
    publisher    = {Zenodo},
    version      = {latest},
    doi          = {10.5281/zenodo.3509134},
    url          = {https://doi.org/10.5281/zenodo.3509134}
}

@inproceedings{Bai:2017qf,
	Adsurl = {http://adsabs.harvard.edu/abs/2017ASSL..445...63B},
	Author = {{Bai}, X.-N.},
	Booktitle = {Astrophysics and Space Science Library},
	Doi = {10.1007/978-3-319-60609-5_3},
	Editor = {{Pessah}, M. and {Gressel}, O.},
	Pages = {63},
	Series = {Astrophysics and Space Science Library},
	Title = {{Wind-Driven Global Evolution of Protoplanetary Disks}},
	Volume = 445,
	Year = 2017}

@article{Banerjee:2006lr,
	Adsurl = {http://adsabs.harvard.edu/abs/2006ApJ...641..949B},
	Author = {{Banerjee}, R. and {Pudritz}, R.~E.},
	Doi = {10.1086/500496},
	Eprint = {arXiv:astro-ph/0508374},
	Journal = {\apj},
	Month = apr,
	Pages = {949-960},
	Title = {{Outflows and Jets from Collapsing Magnetized Cloud Cores}},
	Volume = 641,
	Year = 2006}

@article{Shang:2006lr,
	Adsurl = {http://adsabs.harvard.edu/abs/2006ApJ...649..845S},
	Author = {{Shang}, H. and {Allen}, A. and {Li}, Z.-Y. and {Liu}, C.-F. and {Chou}, M.-Y. and {Anderson}, J.},
	Doi = {10.1086/506513},
	Journal = {\apj},
	Month = oct,
	Pages = {845-855},
	Title = {{A Unified Model for Bipolar Outflows from Young Stars}},
	Volume = 649,
	Year = 2006
}

@INPROCEEDINGS{Shu2000,
       author = {{Shu}, F.~H. and {Najita}, J.~R. and {Shang}, H. and {Li}, Z.-Y.},
        title = "{X-Winds Theory and Observations}",
    booktitle = {Protostars and Planets IV},
         year = 2000,
       editor = {{Mannings}, V. and {Boss}, A.~P. and {Russell}, S.~S.},
        month = may,
        pages = {789-814},
       adsurl = {https://ui.adsabs.harvard.edu/abs/2000prpl.conf..789S}
}

@ARTICLE{Draine2006,
       author = {{Draine}, B.~T.},
        title = "{On the Submillimeter Opacity of Protoplanetary Disks}",
      journal = {\apj},
         year = 2006,
        month = jan,
       volume = {636},
       number = {2},
        pages = {1114-1120},
          doi = {10.1086/498130},
archivePrefix = {arXiv},
       eprint = {astro-ph/0507292},
 primaryClass = {astro-ph},
       adsurl = {https://ui.adsabs.harvard.edu/abs/2006ApJ...636.1114D}
}

@ARTICLE{MRN1977,
       author = {{Mathis}, J.~S. and {Rumpl}, W. and {Nordsieck}, K.~H.},
        title = "{The size distribution of interstellar grains.}",
      journal = {\apj},
         year = 1977,
        month = oct,
       volume = {217},
        pages = {425-433},
          doi = {10.1086/155591},
       adsurl = {https://ui.adsabs.harvard.edu/abs/1977ApJ...217..425M}
}

@ARTICLE{Draine2003a,
       author = {{Draine}, B.~T.},
        title = "{Scattering by Interstellar Dust Grains. I. Optical and Ultraviolet}",
      journal = {\apj},
         year = 2003,
        month = dec,
       volume = {598},
       number = {2},
        pages = {1017-1025},
          doi = {10.1086/379118},
archivePrefix = {arXiv},
       eprint = {astro-ph/0304060},
 primaryClass = {astro-ph},
       adsurl = {https://ui.adsabs.harvard.edu/abs/2003ApJ...598.1017D}
}

@ARTICLE{Draine2003b,
       author = {{Draine}, B.~T.},
        title = "{Scattering by Interstellar Dust Grains. II. X-Rays}",
      journal = {\apj},
         year = 2003,
        month = dec,
       volume = {598},
       number = {2},
        pages = {1026-1037},
          doi = {10.1086/379123},
archivePrefix = {arXiv},
       eprint = {astro-ph/0308251},
 primaryClass = {astro-ph},
       adsurl = {https://ui.adsabs.harvard.edu/abs/2003ApJ...598.1026D}
}

@software{optool2021ascl,
       author = {{Dominik}, Carsten and {Min}, Michiel and {Tazaki}, Ryo},
        title = "{OpTool: Command-line driven tool for creating complex dust opacities}",
 howpublished = {Astrophysics Source Code Library, record ascl:2104.010},
         year = 2021,
        month = apr,
          eid = {ascl:2104.010},
       adsurl = {https://ui.adsabs.harvard.edu/abs/2021ascl.soft04010D}
}

@ARTICLE{OssenkopfHenning1994,
       author = {{Ossenkopf}, V. and {Henning}, Th.},
        title = "{Dust opacities for protostellar cores.}",
      journal = {\aap},
         year = 1994,
        month = nov,
       volume = {291},
        pages = {943-959},
       adsurl = {https://ui.adsabs.harvard.edu/abs/1994A&A...291..943O}
}

@ARTICLE{Andrewsetal2018_DSHARP1,
       author = {{Andrews}, Sean M. and {Huang}, Jane and {P{\'e}rez}, Laura M. and {Isella}, Andrea and {Dullemond}, Cornelis P. and {Kurtovic}, Nicol{\'a}s T. and {Guzm{\'a}n}, Viviana V. and {Carpenter}, John M. and {Wilner}, David J. and {Zhang}, Shangjia and {Zhu}, Zhaohuan and {Birnstiel}, Tilman and {Bai}, Xue-Ning and {Benisty}, Myriam and {Hughes}, A. Meredith and {{\"O}berg}, Karin I. and {Ricci}, Luca},
        title = "{The Disk Substructures at High Angular Resolution Project (DSHARP). I. Motivation, Sample, Calibration, and Overview}",
      journal = {\apjl},
         year = 2018,
        month = dec,
       volume = {869},
       number = {2},
          eid = {L41},
        pages = {L41},
          doi = {10.3847/2041-8213/aaf741},
archivePrefix = {arXiv},
       eprint = {1812.04040},
 primaryClass = {astro-ph.SR},
       adsurl = {https://ui.adsabs.harvard.edu/abs/2018ApJ...869L..41A}
}

@ARTICLE{Pinedaetal2022_PPVII,
       author = {{Pineda}, Jaime E. and {Arzoumanian}, Doris and {Andr{\'e}}, Philippe and {Friesen}, Rachel K. and {Zavagno}, Annie and {Clarke}, Seamus D. and {Inoue}, Tsuyoshi and {Chen}, Che-Yu and {Lee}, Yueh-Ning and {Soler}, Juan D. and {Kuffmeier}, Michael},
        title = "{From Bubbles and Filaments to Cores and Disks: Gas Gathering and Growth of Structure Leading to the Formation of Stellar Systems}",
      journal = {arXiv e-prints},
         year = 2022,
        month = may,
          eid = {arXiv:2205.03935},
        pages = {arXiv:2205.03935},
archivePrefix = {arXiv},
       eprint = {2205.03935},
 primaryClass = {astro-ph.GA},
       adsurl = {https://ui.adsabs.harvard.edu/abs/2022arXiv220503935P}
}

@ARTICLE{benistyetal2022_ppvii,
       author = {{Benisty}, M. and {Dominik}, C. and {Follette}, K. and {Garufi}, A. and {Ginski}, C. and {Hashimoto}, J. and {Keppler}, M. and {Kley}, W. and {Monnier}, J.},
        title = "{Optical and Near-infrared View of Planet-forming Disks and Protoplanets}",
      journal = {arXiv e-prints},
         year = 2022,
        month = mar,
          eid = {arXiv:2203.09991},
        pages = {arXiv:2203.09991},
archivePrefix = {arXiv},
       eprint = {2203.09991},
 primaryClass = {astro-ph.EP},
       adsurl = {https://ui.adsabs.harvard.edu/abs/2022arXiv220309991B}
}
\bibliographystyle{aasjournalv7}

\end{document}